\documentclass{article}

\usepackage{microtype}
\usepackage{graphicx}
\usepackage{makecell}
\usepackage{subfig}
\usepackage{booktabs} 

\usepackage{hyperref}
\newcommand{\angstrom}{\mbox{\normalfont\AA}}

\usepackage{algpseudocode}

\makeatletter
\newenvironment{breakablealgorithm}
  {
   \begin{center}
     \refstepcounter{algorithm}
     \hrule height.8pt depth0pt \kern2pt
     \renewcommand{\caption}[2][\relax]{
       {\raggedright\textbf{\ALG@name~\thealgorithm} ##2\par}%
       \ifx\relax##1\relax 
         \addcontentsline{loa}{algorithm}{\protect\numberline{\thealgorithm}##2}%
       \else 
         \addcontentsline{loa}{algorithm}{\protect\numberline{\thealgorithm}##1}%
       \fi
       \kern2pt\hrule\kern2pt
     }
  }{
     \kern2pt\hrule\relax
   \end{center}
  }
\makeatother

\usepackage[accepted]{icml2023}

\usepackage{amsmath}
\usepackage{amssymb}
\usepackage{mathtools}
\usepackage{amsthm}
\usepackage{bbm}

\usepackage[capitalize,noabbrev]{cleveref}

\theoremstyle{plain}

\theoremstyle{definition}

\theoremstyle{remark}

\usepackage[textsize=tiny]{todonotes}
\usepackage{xspace}
\newcommand{\method}{HierDiff\xspace}

\icmltitlerunning{Coarse-to-Fine: a Hierarchical Diffusion Model for Molecule Generation in 3D}

\begin{document}

\twocolumn[
\icmltitle{Coarse-to-Fine: a Hierarchical Diffusion Model for Molecule Generation in 3D}



\icmlsetsymbol{equal}{*}

\begin{icmlauthorlist}
\icmlauthor{Bo Qiang}{equal,pku}
\icmlauthor{Yuxuan Song}{equal,air}
\icmlauthor{Minkai Xu}{stanford}
\icmlauthor{Jingjing Gong}{air}
\icmlauthor{Bowen Gao}{air}\\
\icmlauthor{Hao Zhou}{air}
\icmlauthor{Weiying Ma}{air}
\icmlauthor{Yanyan Lan}{air}
\end{icmlauthorlist}

\icmlaffiliation{air}{Institute for AI Industry Research (AIR), Tsinghua University}
\icmlaffiliation{pku}{Department of Pharmaceutical Science, Peking University}
\icmlaffiliation{stanford}{Department of Computer Science, Stanford University}

\icmlcorrespondingauthor{Yanyan Lan}{lanyanyan@tsinghua.edu.cn}

\icmlkeywords{Machine Learning, ICML}

\vskip 0.3in
]



\printAffiliationsAndNotice{\icmlEqualContribution.~~ Work was done while Bo Qiang was a research intern at AIR} 

\begin{abstract}
Generating desirable molecular structures in 3D is a fundamental problem for drug discovery. Despite the considerable progress we have achieved, existing methods usually generate molecules in atom resolution and ignore intrinsic local structures such as rings, which leads to poor quality in generated structures, especially when generating large molecules. Fragment-based molecule generation is a promising strategy, however, it is nontrivial to be adapted for 3D non-autoregressive generations because of the combinational optimization problems. In this paper, we utilize a coarse-to-fine strategy to tackle this problem, in which a Hierarchical Diffusion-based model (i.e.~HierDiff) is proposed to preserve the validity of local segments without relying on autoregressive modeling. Specifically, HierDiff first generates coarse-grained molecule geometries via an equivariant diffusion process, where each coarse-grained node reflects a fragment in a molecule. Then the coarse-grained nodes are decoded into fine-grained fragments by a message-passing process and a newly designed iterative refined sampling module. Lastly, the fine-grained fragments are then assembled to derive a complete atomic molecular structure. Extensive experiments demonstrate that HierDiff consistently improves the quality of molecule generation over existing methods\footnote{Code is available at \url{https://github.com/qiangbo1222/HierDiff}}.
\end{abstract}

\section{introduction}
Deep generative models have specifically shown great promise in modeling complex graph-like molecular structures, ranging from generating molecular atom-bond graphs~\citep{li2018learning, liu2018constrained, jin2018junction} to generating molecular conformations from graphs~\citep{xu2022geodiff, torsion_diff}. Despite the significant progress achieved, a remaining but vital research direction in this track is de novo design of drug molecules in 3D space. Integrating the 3D information into the molecule design process enjoys several advantages over only involving topological information in many important applications, \emph{e.g.}, structure-based drug design~\citep{23iclr_pocketfrag, luo2022pocket2mol, luo20213d}, molecular dynamic simulation~\citep{HANSSON2002190}, 3D similarity searching~\citep{shin2015three}. 

Some early studies on 3D molecule generation usually adopt an autoregressive approach~\citep{Gschnet, luo20213d, li2021structure}, which introduces an artificial order on the atoms and generates the atoms one by one in a language generation way. However, molecules have a natural geometric structure in 3D. Besides, these models suffer from the scale and error accumulation problem~\citep{shaw2022generating, G-spherenet}. To tackle these problems, non-autoregressive models have been introduced in this area and gain impressive results. For example, inspired by the successful diffusion model in text and image generation, \citet{edm} proposed the first diffusion model for molecule generation and significantly improves the validity of generated molecules. 

\begin{figure}[h]
\begin{center}

\includegraphics[width=0.35\textwidth]{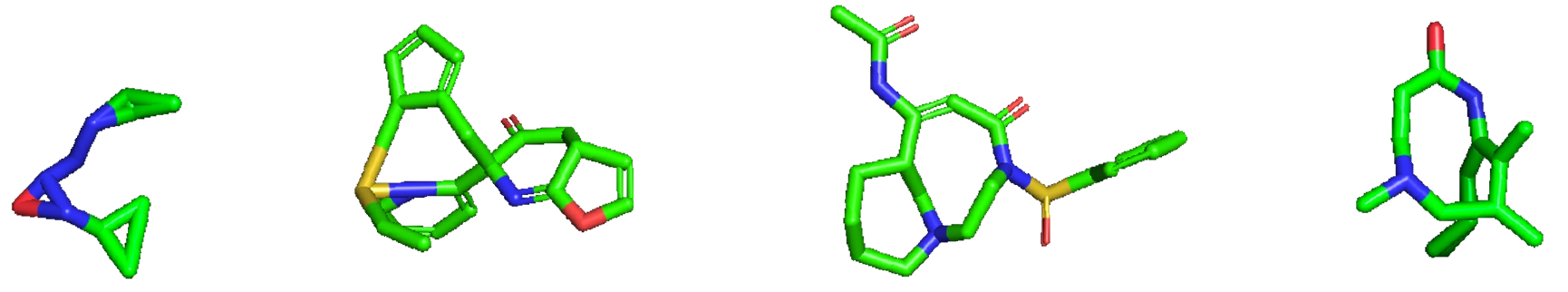}
\end{center}
\vspace{-15pt}
\caption{Visualization results of 3D conformations generated by atom-based methods.}
\vspace{-1pt}
\label{fig:atom-vis}

\end{figure}

Nevertheless, the atom-level generation manner in these works, though enjoys higher flexibility to place each atom, lacks the necessary constraints to obtain reliable molecule structures, especially when generating large molecules. As shown in Figure~\ref{fig:atom-vis}, without imposing Euclidean geometric constraints on the modeling process, the generated 3D aromatic rings could seriously violate basic chemical rules. 

In this paper, we propose a coarse-to-fine approach to tackle the above problems. The basic idea is that we first generate 
the coarse-grained structure of the molecule, where each node represents a cluster of fragments, and then the coarse-grained structure is decoded into fine-grained fragments to assemble atomic molecule structure. In this way, valid local structures will be preserved by replacing the computation unit from atoms with fragments. However, such a process is non-trivial since generated neighborhood fragments may suffer from atom-bonds conflicts, preventing them to be connected. 

To tackle the problem, we treat 3D molecule generation as a constraint generation problem and propose a novel Hierarchical Diffusion-based model (HierDiff). In the coarse-grained phase, our method generates the \textit{fragment representation} instead of \textit{deterministic fragment}. Specifically, we introduce two different ways to obtain chemically interpretable features for representing fragments. Then we propose a geometric diffusion model to generate these fragment representations and their Cartesian coordinates in an efficient non-autoregressive manner. In the fine-grained phase, we utilize an equivariant message-passing network to guarantee connectivity and an iterative refinement module to correct the bias. At last, we construct atom-level 3D structure based on the decoded 3D fragment graph. 

The proposed coarse-to-fine approach nicely mimics the chemistry expert’s drug design process by combining fragments from a pre-defined functional group database. In this way, important inductive biases in this area are encoded in our model.
Furthermore, from machine learning’s perspective, the fragment-based representation of molecules significantly reduces unnecessary degrees of freedom in the atom-based methods,
thus will lead to global optimum convergence and better generalization ability. 

Extensive experiments are conducted to test our model on the challenging task of generating drug-size molecules. Compared to the baseline model, HierDiff can generate both realistic molecules with better drug-like properties and conformations that are much closer to the ground truth conformations. Visualized results also demonstrate HierDiff is capable of generating high-quality molecules with more stable substructures.

\section{Related work}
Molecule generation is one of the fundamental problems in drug discovery. In earlier works, molecule generation tasks are tackled by generating sequential representations of molecules, i.e.~SMILES \textit{e.g.} ~\citep{kusner2017grammar, dai2018syntax, segler2018generating}. With the development of graph neural networks, researchers begin to utilize the graph-based generative model to generate molecular topological structures and gain great progress~\citep{jin2018learning, jin2018junction, li2018multi} However, neither sequence nor graph-based models capture the 3D geometric information, which is crucial for various molecule applications, such as molecule property prediction and protein-ligand docking. Recently, 3D molecule generation has become an emerging hot topic in this area, and different deep generative models have been proposed to tackle this problem. For example, Gschnet~\citep{Gschnet} employs an autoregressive process equipped with Schnet~\citep{Schnet} to sample atoms and bonds iteratively. G-spherenet~\cite{G-spherenet} applied discrete flows to autoregressively generate invariant geometric features. EnFlow~\citep{enflow} utilizes continuous time normalizing flows to sample valid molecules. EDM~\citep{edm} is the first to apply the powerful diffusion model to this area and gains further improvements. However, EDM always generates unrealistic ring systems and broken molecules, when training on large molecules. 

A related branch of molecule generation is hierarchical graph generation. Most previous works derive the hierarchical structure based on some intrinsic rules. For example, some work uses different granularity levels, such as atom-motif \citep{jin2020hierarchical,jin2019hierarchical}, or node-edge\citep{xianduo2022hierarchical}, to construct different hierarchies. \citep{zhou2019misc} and \cite{chauhan2019multiscale} use predefined rules to distinguish different nodes to different levels. \citep{mi2021hdmapgen} employs the natural graphical topology to define the hierarchy. \citep{23iclr_motifmine} collects connection information to form a hierarchical structure. \citep{kuznetsov2021molgrow} obtain the hierarchy by adding latent variables to different layers of the model. While in our method, we use the learnable decoding module to approximate a semantic-guided hierarchy.

\section{Backgrounds}
\subsection{Denoising Diffusion Probabilistic Model}
\label{sec:ddpm}
 Denoising diffusion probabilistic model (DDPM)~\citep{DDPM, diff_prob_model} provides a powerful generative modeling tool by reversing a diffusion process. More specifically, the \emph{diffusion} process projects the noise into the ground truth data and the \emph{generative} process learns to reverse the process. The two processes imply a latent variable model, where $\mathbf{x}_1,\cdots ,\mathbf{x}_{t-1}$ are the latent variables. The \emph{forward} process could be seen as a fixed approximate posterior distribution:
     \begin{equation}
         \begin{aligned}
         \label{0}
         q_(\mathbf{x}_{1:T}|\mathbf{x}_0) &= \prod_{t=1}^{T} q(\mathbf{x}_t|\mathbf{x}_{t-1})\\
         q\left(\mathbf{x}_t \mid \mathbf{x}_{t-1}\right)&=\mathcal{N}\left(\mathbf{x}_t; \sqrt{1-\beta_t} \mathbf{x}_{t-1}, \beta_t \mathbf{I}\right)
         \end{aligned}
     \end{equation}

 Here $\beta_1,\cdots,\beta_T$ corresponds to a fixed variance schedule. For simplicity, we let $\alpha_t=1-\beta_t$ and $\bar{\alpha}_t=\prod_{i=1}^t \alpha_i$, the forward pass for arbitrary time step has an analytic form, \emph{i.e.}, $q\left(\mathbf{x}_t \mid \mathbf{x}_0\right)=\mathcal{N}\left(\mathbf{x}_t ; \sqrt{\bar{\alpha}_t} \mathbf{x}_0,\left(1-\bar{\alpha}_t\right) \mathbf{I}\right)$. The \emph{generative} process parameterized the transition kernel $P_\theta(\mathbf{x}_{t-1}|\mathbf{x}_{t})$ of the Markov chains, the corresponding likelihood function could be derived as:
     \begin{equation}
         \begin{aligned}
     \label{eq:generative_diffusion}
         P_\theta\left(\mathbf{x}_{t-1} \mid \mathbf{x}_t\right)&=\mathcal{N}\left(\mathbf{x}_{t-1} ; \mu_\theta\left(\mathbf{x}_t, t\right), \sigma_t^2 \mathbf{I}\right)\\
         P_\theta(\mathbf{x}_0)&=\int p\left(\mathbf{x}_T\right) P_\theta\left(\mathbf{x}_{0: T-1} \mid  \mathbf{x}_T\right) \mathrm{d} \mathbf{x}_{1: T}
         \end{aligned}
     \end{equation}
 Here the $\mu_\theta$ refers to parameterized means function and the $\sigma_t^2$ is the predefined variance. For the initial distribution $P_\theta(\mathbf{x}_T)$, we select invariant base distribution for equivariant coordinates.

 \subsection{Equivariance and SE(3)-invariant Density Estimation}
 \label{sec:eqv}
 Equivariance widely exists in the physical world, especially in atomic systems. For example, the vector fields of atomic forces should rotate or translate correspondingly with the 3D positions of the molecule. Thus integrating such inductive bias into the function modeling has appealing properties and has been widely explored~\citep{pointconv, Schnet, EGNN}. More specifically, given two transformation $T_g$ and $S_g$ acting on the space $\mathcal{X}$ and $\mathcal{Y}$, a function $f$ is considered as equivariant with respect to the group $G$ if the following is satisfied:
 {\setlength\abovedisplayskip{+0.8pt}
 \setlength\belowdisplayskip{+0.8pt}
 \begin{align}
     f \circ T_g(x)=S_g \circ f(x)
 \end{align}
 }
 In this task, we mainly focus on the SE(3) group, \emph{i.e.}, the group of rotation and translation in the 3D space. 
 For generative modeling of 3D molecule graph, the density function of the model distribution $P_\theta(.)$ should be SE(3)-invariant, \emph{i.e.}, $P_\theta(x) = P_\theta(T_g(x))$.
 To this end, previous methods either directly model the invariant components, \emph{e.g.}, bond angles, or use some invariant base distribution and model the transformation by the equivariant neural network~\citep{torsion_diff, enflow}. \method extends the latter one to an equivariant hierarchical framework to model fragment coordinates and fit bond lengths to predefined rules, which will be discussed in Sec.~\ref{sec:method}.
 
\begin{figure}[t]
 \centering
 \subfloat[Conflict Free]{\includegraphics[width=0.12\textwidth]{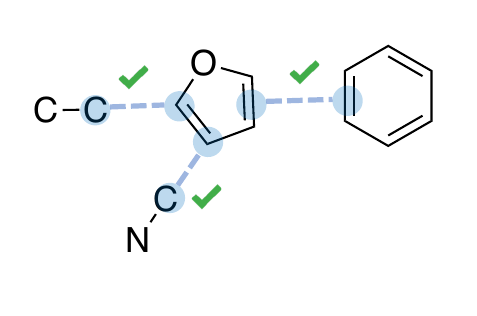}}
 \hspace{1em}
 \subfloat[One-hop]{\includegraphics[width=0.14\textwidth]{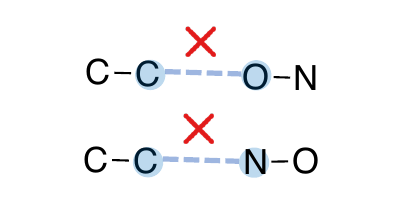}}
 \hspace{1em}
 \subfloat[Two-hop]{\includegraphics[width=0.14\textwidth]{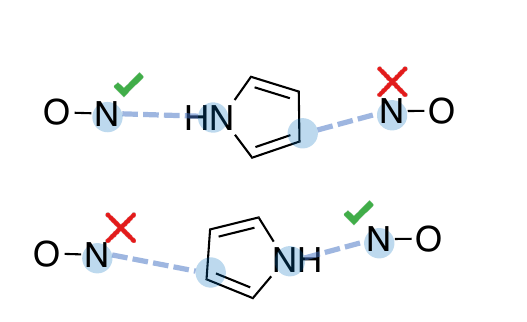}}
  \caption{An illustration of fragment conflicts. One-hop conflict means the two linked fragments do not share any elements to form a valid edge. Two-hop conflict represents that though linked fragments can form edges by sharing the same atom/bond, conflicts occur when the valence is violated}\label{appd:cfl}
  \vspace{-8pt}
  \label{fig:conflict}
\end{figure}

\section{Coarse-to-Fine Approach}

In this section, we introduce our coarse-to-fine approach, including problem formulation and coarse-to-fine definition.

\subsection{Problem Formulation and Notations}

Let $\mathcal{G}$ be the space of 3D graphs, where each 3D graph $G$ consists of the fragment set $\mathcal{V}$ and the edge set $\mathcal{E}$. More specifically, every fragment $V\in  \mathcal{V}$ represents a combination of several atoms and bonds, \emph{e.g.}, a benzene ring could be a fragment that includes six carbon atoms and aromatic bonds. Instead of using edges to represent chemical bonds as in previous works, we use edge $ E_{ij} \in \mathcal{E}$ to indicate that there is a bond/atom  shared by two fragments $V_i$ and $V_j$. This kind of definition enables us to model molecule geometry using tangent condition on fragment sphere, as Fig.~\ref{fig:tangent}
, with a reasonable size fragment vocabulary. Therefore, the target of a 3D generation model is to learn a probabilistic model $P_\theta(\mathcal{V},\mathcal{E})$ to model the empirical distribution of 3D molecule graphs, which could also be used to sample new molecules.

\begin{figure*}[t]

\begin{center}
\includegraphics[width=0.8\textwidth]{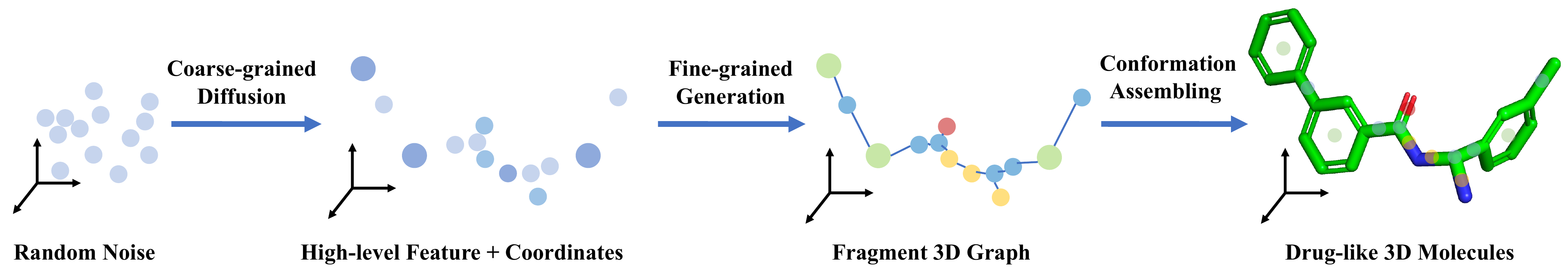}
\end{center}
\vspace{-10pt}
\caption{An overview of the hierarchical diffusion model. }
\label{fig:framework}
\vspace{-10pt}
\end{figure*}

\subsection{Coarse-to-Fine Framework}
\label{challenge}
\subsubsection{Challenges of Non-autoregressive Methods}
 
Compared with the autoregressive approach, non-autoregressive generative models are more promising for 3D molecule generation, due to their natural advantages of global modeling ability~\citep{de2018molgan, kwon2019efficient, enflow}. Empirically, previous work has shown consistent observations~\citep{edm}.

Though there are several appealing properties, the non-autoregressive model at the fragment level indeed implies the following structure generation procedure under hard constraints.
\begin{equation}
\label{eq:constrain_gen}
    \begin{aligned}
        &\,\,G \sim P_\theta(\mathcal{V},\mathcal{E}),\\
        &s.t.~G_s(V_i) \in \mathcal{W},~\forall i=1,\cdots n,
    \end{aligned}
\end{equation}
where $G_s(V_i)$ stands for the substructure which consists of $V_i$ and its neighbors, $\mathcal{W}$ stands for the set of all valid substructures.
Intuitively, the valid substructures satisfy the following conditions: the neighbors should hold matched components(atoms/bonds) to get ensembled; for the cases where a node has multiple neighbors there should also be enough matched components in this node to match all its neighbors. We provide Fig.~\ref{appd:cfl} to better illustrate the constraints in Eq.~(\ref{eq:constrain_gen}). 

Limited chemical valency makes conflicts really common in fragment generation. Note that the problem of avoiding fragment conflict has high complexity and brings the so-called "combinatorial exploding" issues\citep{jin2020hierarchical} because of the multi-hop conflicts. For non-autoregressive modeling fashion, the complexity increases exponentially with the structure size. This problem is also discovered in other tasks, for example, Dispatching Route Generation~\citep{ding2021xor}, Optimal Experiment Design~\citep{le2012streamlined}, and  Protein Alignment Generation~\citep{xu2015protein}. In fragment-based molecule generation, it is challenging to generate realistic drug-size molecules in an ordering-agnostic way. 
\subsubsection{Solutions for Avoiding Conflicts}

It is difficult to sample from the distribution with hard constraints.
One direct solution that was adopted in previous work~\citep{popova2019molecularrnn} is to conduct rejection sampling, \emph{i.e.}, only accept the connectable molecules. Nevertheless, rejection sampling is not applicable in practice for fragment-based methods due to the extremely low acceptance rate; Alternatively, one can use a learnable model, $P_\phi(V_i|N(V_i))$, to approximate the hard constraint, where $N(V_i)$ stands for the neighbors of $V_i$. Correspondingly, the generative distribution $P_{\theta,\phi}(\mathcal{V},\mathcal{E})$ could be written as another target distribution $ P_\theta(\mathcal{V},\mathcal{E}) \prod_{1\leq i \leq n}P_\phi(V_i|N(V_i))$. Unfortunately, Markov chain Monte Carlo (MCMC) sampling is needed, such as Gibbs Sampling, to conduct sampling from such distribution, which still suffers from efficiency issues.

Instead of making generated samples satisfy the constraint through filter or refinement, we try to decompose and embed the constraint directly into the  model phase within a hierarchical fashion.

More specifically,  
we design the variable~($H$) as the latent variable and the probabilistic model could be expressed as $P_{\theta,\phi}(\mathcal{V},\mathcal{E}) = P_\theta(H) P_\phi(\mathcal{V},\mathcal{E}|H)$.

Consequently, we could obtain a lower bound of the maximum likelihood objective by the concavity of the logarithm, as follows:
\begin{equation}\label{eq:abs_objective}
\small
\begin{aligned}
\mathbb{E}_{(V,E) \sim P_{\text{data}}} \log \sum_{H \in \mathbb{H}} & P_\theta(H) P_\phi(V,E|H) \geq  \\ 
\mathbb{E}_{(V,E) \sim P_{\text{data}}} & \mathbb{E}_{H\sim Q(H|V,E)} \Bigg[\underbrace{\log P_\theta(H)}_{\text{Coarse-grained Diffusion}} \\ 
& +\underbrace{\log P_\phi(V,E|H)}_{\text{Fine-Grained Generation}} -\underbrace{\log Q(H|V,E)}_{\text{Constant Term}} \Bigg] 
\end{aligned}
\end{equation}
where $\mathbb{H}$ stands for the possible support of $H$. Formally, $Q(H|V,E)$ in Eq.~(\ref{eq:abs_objective}) is implemented by extracting chemical features and averaging the atom coordinates. As $Q(H|\mathcal{V}, \mathcal{E})$ is set to a constant term, the objective will include only the former two terms, i.e.~$\log P_\theta(H)$ and $\log P_\phi(\mathcal{V},\mathcal{E}|H)$. The first objective $\log P_\theta(H)$ could be approximated by a coarse-grained fragment diffusion model and the second objective $\log P_\phi(\mathcal{V},\mathcal{E}|H)$ is modeled by equivariant message passing networks and iterative refinement process. 

Intuitively, the coarse-to-fine process is like first determining the position and the function of each component, then finding the connectable fragments from small subsets and assembling them. Therefore, \method could maintain the global modeling property of non-autoregressive methods and also significantly reduce the complexity of finding the connectable fragments.

\section{HierDiff: hierarchical diffusion-based model}
\label{sec:method}
In this section, we introduce the proposed \method model in detail, as illustrated in Fig.~\ref{fig:framework}, including coarse-grained fragment generation, fine-grained fragment generation, and atom conformation assembling which also correspond to the parameterized terms in Eq.~(\ref{eq:abs_objective}). 

\subsection{Coarse-Grained Fragment Diffusion}
\label{sec:coarse_set}

We define $H = [H_f, H_p]$ as the representation of the coarse nodes, where $H_f$ stands for the invariant chemical features and $H_p$ stands for the equivariant positional features. Formally, $Q(H|V,E)$ in Eq.~(\ref{eq:abs_objective}) is implemented by extracting chemical features to obtain $H_f$ and averaging all the atom coordinates to obtain $H_p$. Specifically, the property-based features could depend on both fragment $V$ and the attachment $E$, i.e.~the connection to neighbor fragments. 

In the coarse-grained phase, a diffusion model is proposed to approximate $\log P_\theta(H)$. Note that, when sampling from the diffusion model, we first sample the number of coarse nodes from the histogram we calculated on the training set.

\subsubsection{Chemical Feature}

We carefully design the features to be discriminative enough for fragments and molecules with different chemical and geometrical properties, which allows us to easily integrate our domain knowledge as inductive bias into the model. And we specifically employ two kinds of features:

\textbf{Property-based Coarse Feature}: We summarize some important properties which are widely used in drug discovery into an 8-dimension vector, including the number of hydrogen bonds and rings, the area of different surfaces, etc. \\

\textbf{Element-based Coarse Feature}: 
We also include the histogram of element frequency, \emph{i.e.}, a 3-dimension vector, to the feature representation, inspired by the fact that elements with the same number of valence electrons usually share the same properties. 

\begin{figure}[t]
\begin{center}
\includegraphics[width=0.4\textwidth]{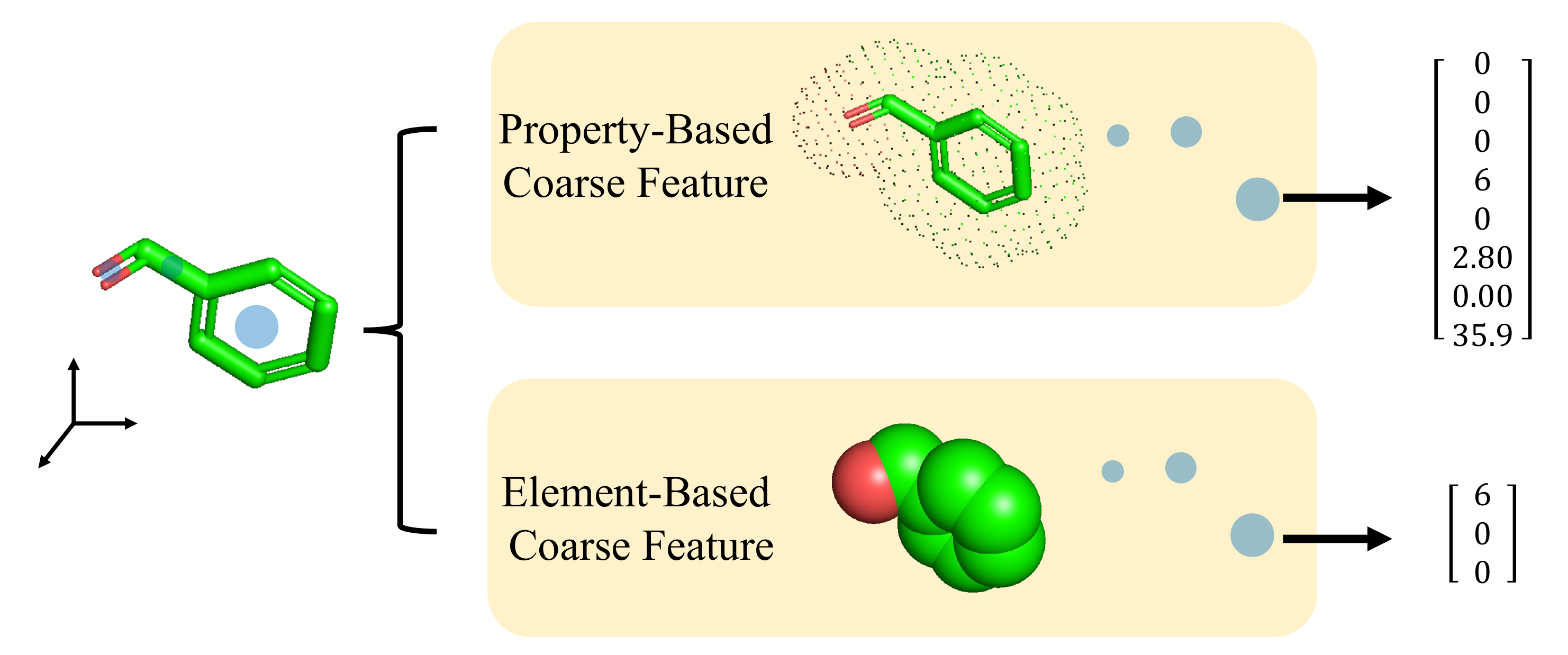}
\end{center}
\vspace{-10pt}
\caption{Illustration of the featurization of 3D Benzaldehyde using two kinds of features.}
\label{fig:feat_vis}
\end{figure}

Please refer to Fig.~\ref{fig:feat_vis}, Table~\ref{tab:property_based} and Table~\ref{tab:element_based} for the detailed implementations. 

\subsubsection{Positional Feature}

\begin{figure}[t]
\begin{center}
\includegraphics[width=0.4\textwidth]{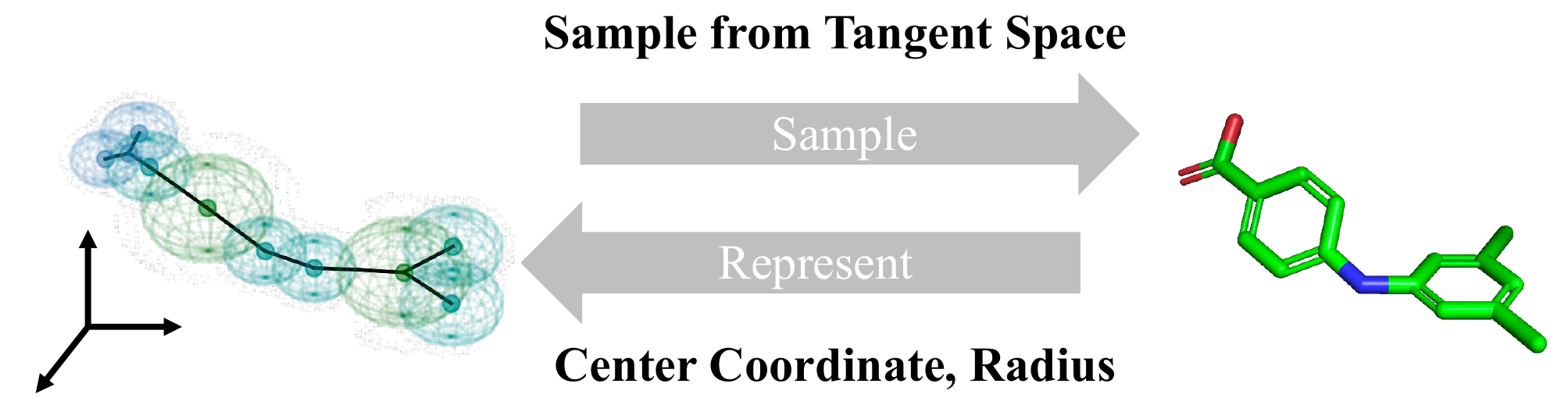}
\end{center}

\caption{The spheres of all coarse-grained nodes will be determined since the radius of coarse-grained nodes could be directly determined by basic chemical rules. Conversely, the full atom conformation can be reconstructed, by sampling positions on the tangent of intersecting spheres. Therefore, the center coordinate is a good representation.}
\label{fig:tangent}
\end{figure}

There are several possible ways to represent the 3D conformation systems in fragment level, \emph{e.g.}, the dihedral angle between neighbor fragments, and the distance matrix. In this paper, we simply use the center coordinates as the positional feature of the coarse node, since we found that this information is enough to determine the conformation at atom resolution, with a predefined vocabulary of bond length and bond angles from the RDkit ETKDG module. Please note that the center position of the coarse node could be seen as the center of the conformation sphere which includes all possible conformations generated from the degree of freedom on rotation. The connected fragments correspond to the tangent condition which actually eliminates the degree of freedom, which is illustrated in Fig.~\ref{fig:tangent}.

\subsubsection{Diffusion Process}
\label{sec:diffpro}

Then, we introduce the modeling of $H_f$ and $H_p$ individually.

$H_f$ could be modeled by a typical diffusion model with Gaussian noise for step $t > 0$. However, we find that the $0$-th term for continuous feature $H_f^{\text{int}}$ and $H_f^{\text{cont}}$, \emph{i.e.} $\mathcal{L}_0$, should be designed carefully, as observed similarly in~\citep{edm}. In this paper, we use the following form, which has shown a better empirical performance.
\begin{equation*}
\begin{aligned}
    \mathcal{L}_0(H_f^{\text{int}},H_f^{\text{cont}})= - \log [\int_{H_f^{\text{int}}-\frac{1}{2}}^{H_f^{\text{int}} + \frac { 1 } { 2 }} \mathcal{N}\left(\boldsymbol{u} \mid \boldsymbol{x}_0^{(H_f^{\text{int}})},\sigma_0\right) \mathrm{d} \boldsymbol{u}] \\
    -  \log\mathcal{N}\left(H_f^{\text{cont}}\mid \frac{\boldsymbol{x}_0^{(H_f^{\text{cont}})}} {\alpha_0}- \frac{\sigma_0} {\alpha_0} \hat{\boldsymbol{\epsilon}}_0, \frac{\sigma_0^2}{\alpha_0^2}  \mathbf{I}\right).
\end{aligned}
\end{equation*}

Next, we describe the generation for $H_p$.

To make the likelihood function in Eq.~(\ref{eq:generative_diffusion}) to be SE(3)-invariant, we set the initial distribution under the zero center of mass~(CoM) systems~\citep{kohler2020equivariant}, \emph{i.e.}, applying a CoM-free Gaussian:
\begin{align*}
    \label{Eq:initial_distribution}
    \mathcal{N}\left(H_p \mid \boldsymbol{0}, \sigma^2 \mathbf{I}\right)=(\sqrt{2 \pi} \sigma)^{-(M-1) \cdot n} \exp \left(-\frac{1}{2 \sigma^2}\|H_p\|^2\right).
\end{align*}
Here $H_p$ belongs to the space $\mathbb{R}^{M \times n}$, where $M$ is the number of fragment nodes and $n$ equals the coordinate dimension. Besides, an equivariant Markov transition kernel is constructed under the widely applied noise parameterization~\citep{ho2020denoising}:
\begin{align*}
    \mu_\theta\left(H_{p}^t, t\right)=\frac{1}{\sqrt{\alpha_t}}\left({H}_{p}^t-\frac{\beta_t}{\sqrt{1-\bar{\alpha}_t}} \epsilon_\theta\left(H_{p}^t, t\right)\right).
\end{align*}

If $\epsilon_\theta$ is parameterized by SE(3)-equivariant networks, the transitional kernel $P_\theta(H_{p}^{t-1}|H_{p}^{t-1})$ is also SE(3)-equivariant, \emph{i.e.}, $P_\theta \left(H_{p}^{t-1} \mid H_{p}^{t}\right)=P_\theta \left(T_g\left(H_{p}^{t-1} \right) \mid T_g\left(H_{p}^{t}\right)\right)$~\citep{xu2022geodiff}. We leave the detailed proof in Appendix~\ref{appd:ddpm}.

\subsection{Fine-Grained Fragment Generation} 
\label{sec:frag assembl}
    \begin{figure*}[t]
\begin{center}
    
\includegraphics[width=1\textwidth]{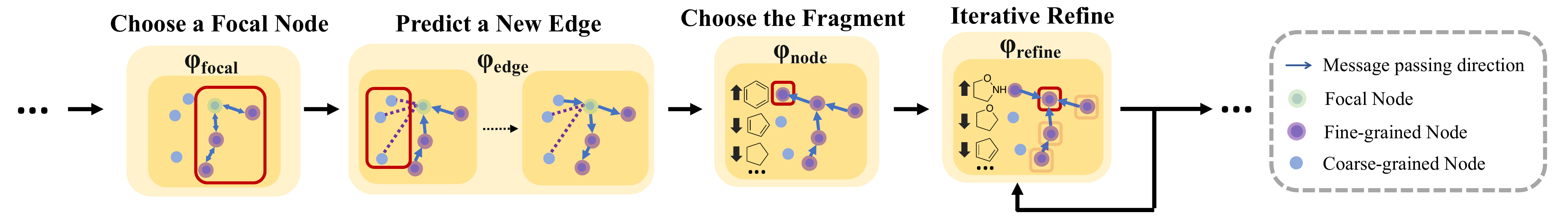}
\vspace{-1pt}
\caption{Illustration of the fine-grained atom generation process, including (1)~choosing a focal node among fine-grained nodes using \textbf{vanilla EGNN}, (2)~predicting the link from all candidate new edges marked with dashed lines using \textbf{bottom-up or top-down EGNN}, (3)~generating the exact fragment type using \textbf{bottom-up EGNN}, (4)~iterative refining the previous fine-grained node via sampling from distribution parameterized by \textbf{bottom-up EGNN}.} 
\vspace{-8pt}
\label{fig:sample}
\end{center}
\end{figure*}

After $[H_f, H_p]$ were generated by the diffusion model, a set of coarse-grained nodes in 3D space is obtained, illustrated in Fig.~\ref{fig:framework}. Now we introduce the detailed process of generating fine-grained fragment types and edges, conditioned on the coarse-grained nodes, which correspond to the term $P_\phi(\mathcal{V}, \mathcal{E}|H)$.

We first briefly introduce the decoding logic here. In each decoding step, the fine-grained generation process contains four stages. Firstly, we select a focal node from all the existing fine-grained nodes with a parameterized neural network module $\phi_{\text{focal}}$. Next, we utilize a link prediction network, \emph{i.e.} $\phi_{\text{edge}}$, to identify a new node that could be linked to the focal node from all the remaining coarse-grained nodes. Then we obtain the fine-grained fragment type of the above newly linked coarse-grained node with the help of another network $\phi_{\text{node}}$. At last, an iterative refinement process is conducted to correct the bias in fine-grained nodes, based on the newly determined fragment type. It should be noted that in the beginning, all the nodes are coarse-grained, so we randomly select a node and directly use $\phi_{\text{node}}$ to predict its fragment type. The above procedure is illustrated in Fig.~\ref{fig:sample}. We emphasize several key elements of our assembling module
in the following:

\subsubsection{Iterative Refinement}
Now we delve into the details of the iterative refinement process. The motivation for introducing this iterative refinement process is to correct the bias in the existing fine-grained nodes, to enhance the ability to generate more realistic molecules from a global view. Specifically, we design a mask prediction model $\phi_{\text{refine}}$ to approximate the probability of each decoded fine-grained fragment conditioned on all coarse-grained nodes and the other fine-grained nodes. The target is to maximize the joint probability of fine-grained nodes as follows:
\begin{equation}
         \begin{aligned}
\label{6}
P_{\text{target}} = \prod_{V_i \in T_f}  P_{\phi_{\text{refine}}}(f(V_i) \mid T_c, T_f \setminus V_i),
        \end{aligned}
\end{equation}
where $T_f$ is the set of all existing fine-grained nodes, $T_c$ is the set of coarse-grained nodes, and $f$ is a function to return the fragment type for a specific node. To sample from the above target distribution, we defined a Markov Chain, in which node type replacement is defined as the state transition, and an early-stopping Monte Carlo sampling strategy is adopted to conduct the sampling process. The detailed algorithm is illustrated in Appendix~\ref{appd:decode}. We also conduct an ablation study to prove the effectiveness of the proposed iterative refinement process, as shown in Appendix~\ref{appd:abl}.

\subsubsection{Message Passing Neural Networks}

Here we introduce the aforementioned models in the fine-grained process, i.e.~$\phi_{\text{focal}} $, $\phi_{\text{edge}}$, $\phi_{\text{node}}$ and $\phi_{\text{refine}}$. In order to avoid the disconnectivity problem as Fig.\ref{fig:conflict}, we need to elaborately design these models.  

Specifically, the input is a set of 3D fragments, represented by both chemical and positional features. In addition, we add one-hot vectors to indicate the fragment types for all fine-grained nodes. At the initial stage, the input is treated as a fully connected 3D graph and a vanilla EGNN~\citep{EGNN} extracts the initial embeddings for all links and nodes. According to the fine-grained generation process, $\phi_{\text{focal}}$ simply passes information among fine-grained nodes. $\phi_{\text{edge}} $ then aggregates information of all fine-grained nodes to the focal node by a tree bottom-up pattern, in which the focal node is treated as the root of the tree structure. After the new edge is predicted, the network broadcasts the addition of the new edge to all fine-grained nodes in a tree top-down pattern. Finally, $\phi_{\text{node}}$ aggregates the information from all fine-grained nodes in the bottom-up pattern to the new node for decoding the fine-grained fragment type. For mask prediction module $\phi_{\text{refine}}$, we utilize a bottom-up EGNN similar to $\phi_{\text{node}}$. The information is aggregated from all fine-grained nodes to the masked nodes, to compute the target distribution as in Eq.~(\ref{6}). The illustration of the message-passing process could be found in Fig.~\ref{fig:sample}. We discussed model-level modification of EGNN in Appendix~\ref{appd:EGNN}.

\subsubsection{Training}
During training, we start by first randomly sampling a connected subgraph at each step. Then a random leaf node is picked, and we simulate the fine-grained generation of this node
All the fine-grained nodes and edges of the subgraph except the selected one are kept. For the other nodes, we only maintain the coarse features and their position. Then $\phi_{{\text{focal}}}$ is trained based on the above feature to maximize the probability of the parent of the selected node among the nodes in the fine-grained subgraph. $\phi_{\text{edge}}$ is trained to maximize the probability of the edge link between the focal node and the selected node among all other coarse-grained nodes. $\phi_{\text{node}}$ is trained to output the fine-grained fragment type of the selected node. For the iterative refinement part, we just randomly mask a node's fine-grained feature on the subgraph, and $\phi_{\text{refine}}$ is trained to reconstruct its masked fragment type. Detailed implementations and objectives could be found in Appendix~\ref{appd:decode}.

\subsection{Assembling to Atom Conformation}

\label{sec:f2a}
Given all fine-grained nodes and link relations determined in the fine-grained generation process, we have to decide which atoms within two linked fragments could be merged to construct the atom-level conformation. To conduct this process, we first randomly choose a fragment, enumerate all possible attachments for its neighbor fragments, and select the one that has the closest fragment center geometric as our generated positional features in the coarse-grained fragment generation. Specifically, we use RDkit to generate the local conformation for each candidate attachment following~\citep{wangregularized} and apply the root-mean-square deviation (RMSD) to measure the difference between fragment center coordinates. The above process will be continued following the neighboring structure until all the local connections are determined. Then we generate the coordinates of each atom. To plug the local conformations into each molecule coordinate system, we need to determine the rotation matrix ($R$) and translation vector ($t$). Here we compute $R$ and $t$ between generated local coordinates and RDkit predicted local coordinates using Kabsch Algorithm~\citep{kabsch1976solution} at the fragment level. Then we applied the obtained $R$, $t$ on the RDkit generated atom level coordinates to align the RDkit generated local geometry to our sampled center positions. This process starts from the subgraph constructed by a randomly selected fragment and its neighbors and is conducted successively until the full atom conformation is derived. Noted that, RDkit is only utilized for generating local geometry in conformation generation.

The full detailed algorithm is introduced in Appendix~\ref{appd:alg-f2a}.

\section{Experiments}
\label{exp section}
In this work, we mainly focus on generating drug-like molecules. So our main experiments are conducted on the dataset of GEOM$_{\text{DRUG}}$~\citep{geom} and CrossDocked2020~\citep{crossdock}. Specifically, GEOM$_{\text{DRUG}}$ includes 304k drug-like molecules, and CrossDocked2020~\citep{crossdock} contains 100k 3D ligand structures extracted from protein-ligand complexes, respectively. As compared with two well-known 3D molecular generation models, i.e.~EDM~\citep{edm} and G-SphereNet~\citep{G-spherenet}, both versions of HierDiff achieve superior results, where \method-E and \method-P denotes the implementation using the element-based feature and property-based feature as representation, respectively. Though our model is not designed for generating small molecules, we also compare HierDiff with several existing models on QM9~\citep{geom}, a popular benchmark for 3D molecule generation evaluation, which the results are shown in Table~\ref{tab:qm9}.

\subsection{Drug-Likeness Evaluation}
The purpose of our proposed generation method is to fabricate molecules that are similar to authentic drug molecules from scratch, thus it is important to measure how drug-likely are those fabricated molecules to true drug molecules. 

\subsubsection{Evaluation Metrics}
Specifically, we mainly measure the drug-likeliness of a molecule from 6 aspects. Quantitative estimate of drug-likeness~(QED), one of the most widely used metrics for virtual screening, is built on a series of carefully selected molecular properties to evaluate drug-likeness. Retrosynthetic accessibility~(RA) is a machine learning based scoring function based on retrosynthesis protocol that evaluates synthetical accessibilities. Medicinal chemistry filter~(MCF) is the rate of sampled molecules that do not contain any undruggable substructures~\citep{brown2019guacamol}. Synthetic accessibility score~(SAS) is a ruled-based scoring function that evaluates the complexity of synthesizing a structure by organic reactions. 
LogP is the octanol-water partition coefficient which is the main factor that determines the distribution of the drug molecules, and $\Delta_{\text{LogP}}$ indicates the difference between the computed LogP and the ground-truth one on the training distribution. Molecular weight~(MW) is a measure of the sum of the atomic weight values of the atoms in a molecule. An ideal model will sample molecules from the same weight distribution as the training set, denoted as ground-truth MW. In practice, $\Delta_{\text{MW}}$ is usually adopted to measure the difference between the computed MW and the ground-truth one.

\subsubsection{Experimental Results} 
\label{exp:res_1}

Table.~\ref{exp:property} shows the experimental results on GEOM$_{\text{DRUG}}$ and CrossDocked2020. From the results, it is obvious that \method significantly outperforms EDM and G-SphereNet in terms of all the concerned evaluation metrics. 

Specifically, the RA comparison results indicate that with all the sub-graphs derived from a predefined vocabulary, \method generates molecules that are easier to synthesize in wet labs, without dangerous substructures. 
Since $\Delta_{\text{MW}}$ shows how close the distribution of generated molecules is to the training data, we can see that G-SphereNet is unable to generate large molecules, while HierDiff generates molecules with more similar size as the training data, as compared with EDM. We speculate that this is mainly due to the error accumulation problem in the autoregressive approach in G-SphereNet, which is proven by our ablation study in Table~\ref{appd:error-accum} of Appendix. Considering the poor performance of G-SphereNet on generating large molecules, we exclude it in the following conformation experiments.

\begin{figure}[t]
\begin{center}
\includegraphics[width=0.48\textwidth]{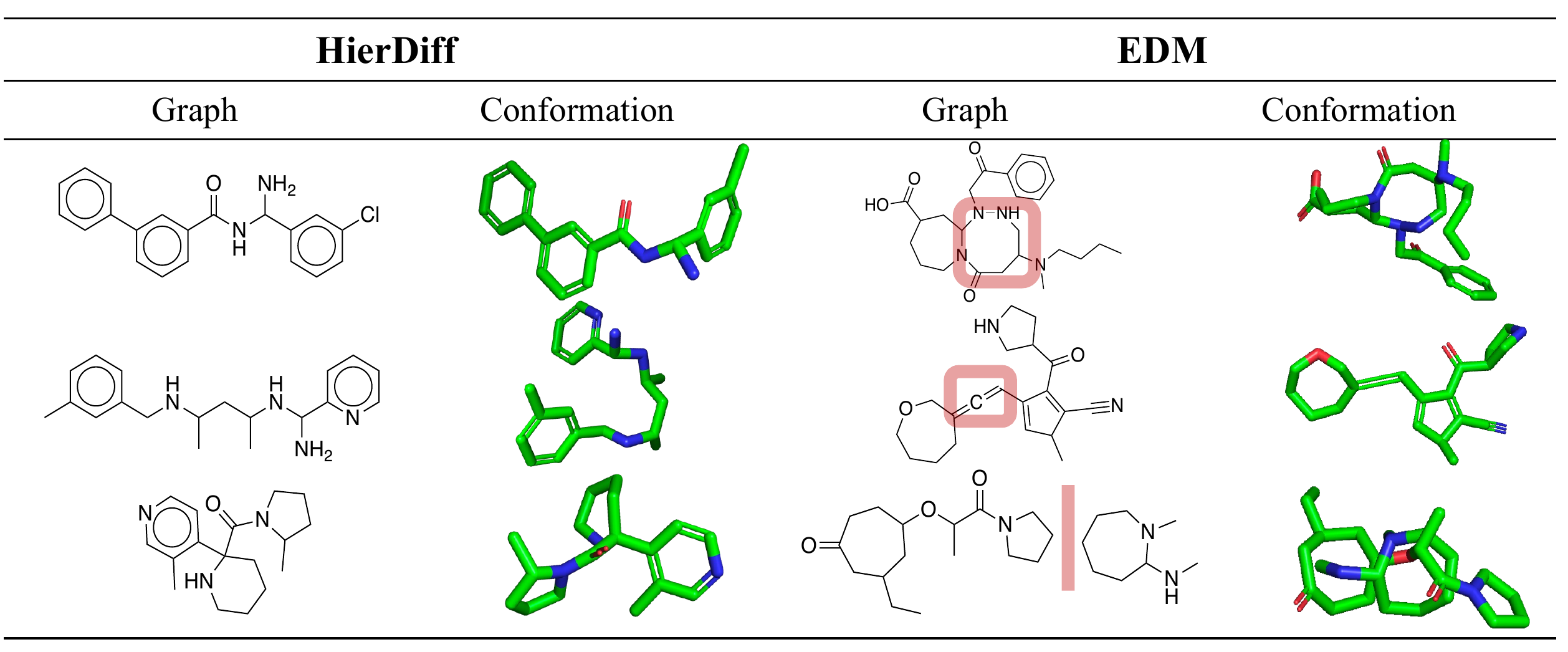}
\end{center}

\caption{Visualization of 3D conformations and 2D molecular graph generated by HierDiff and EDM, where unstable or broken substructures generated by EDM are marked with red boxes or red lines.}

\label{vis_table}
\end{figure}

We also carry out an ablation study on HierDiff w.r.t. the drug-likeliness. Specifically, we remove the iterative refining step and compare the obtained model with the original HierDiff. From the results shown in Appendix~\ref{appd:abl}, the model without iterative refining generates less complicated molecules, as compared with the original HierDiff, in terms of both QED and MCF. Though it has the ability to generate large molecules, such results are far from satisfactory in real applications. Through our analysis, the reason is that, without iterative refinement, HierDiff tends to choose fragments that are easier to assemble, indicating the importance of iterative refinement.

\begin{table}[]
\vspace{-10pt}
\caption{Drug-likeliness evaluation results of the generated molecules on GEOM$_{\text{DRUG}}$ and CrossDocked2020, with the last column standing for the results on training data. GSNet is short for G-SphereNet. Experiments that report errors due to low validity are marked as '--'.}
\label{exp:property}
\begin{center}
\vspace{-5pt}
\resizebox{\linewidth}{!}{
\begin{tabular}{l|cccc|c}
\toprule
 ~ & GSNet & EDM & \method-E & \method-P & \makecell[c]{GEOM\\DRUG}\\
\midrule
QED$\uparrow$  & 0.382 & 0.608 & 0.632 & \bf0.639 & 0.658\\
RA$\uparrow$ & -- & 0.441  & 0.548 & \bf0.639 & 0.915\\
MCF$\uparrow$ & 0.489 & 0.621 &\bf0.727 & 0.659 & 0.774 \\
SAS$\downarrow$ & -- & 4.054 & 3.859 & \bf{3.547} & 4.018 \\
$\Delta_{\text{LogP}}\downarrow$ &2.306 &0.566 & 0.653 &\bf0.128& 0.000 \\
$\Delta_{\text{MW}}\downarrow$ & 170.7& 23.71 & 30.33& \bf13.33& 0.00 \\ 
\bottomrule
\end{tabular}
}
\vspace{+5pt}

\resizebox{\linewidth}{!}{
\begin{tabular}{l|cccc|c}
\toprule
 ~ & GSNet & EDM & \method-E & \method-P & \makecell[c]{Cross\\Dock}\\
\midrule
QED$\uparrow$  & 0.442 & 0.499 & \bf0.614 & 0.585 & 0.619\\
RA$\uparrow$ & -- & 0.332  & \bf0.574 & 0.262 & 0.912\\
MCF$\uparrow$ & 0.449 & 0.613 &\bf0.759 & 0.687 & 0.746 \\
SAS$\downarrow$ & -- & 7.056 & \bf4.051 & 5.397 & 2.564 \\
$\Delta_{\text{LogP}}\downarrow$ &3.359 &2.840 & 1.872 &\bf1.176& 0.000 \\
$\Delta_{\text{MW}}\downarrow$ & 200.95& 28.40 & 44.56& \bf1.15& 0.00 \\
\bottomrule
\end{tabular}
}
\vspace{-15pt}
\end{center}
\end{table}

\subsection{Conformation Quality Evaluation}
\label{exp:conf}


\subsubsection{Evaluation Metrics}
However, since all generated molecules are new, i.e.~not existing in training data, we cannot directly use ground-truth conformations from the database for evaluation. To obtain corresponding ground-truth conformation, we adopted the same experimental procedure as in~\cite{geom}. which has been proven to be suitable for evaluating the conformation quality 3D conformation generation in several previous work~\cite{xu2022geodiff, torsion_diff}. Specifically, a computational costly molecular dynamic simulation~(detailed experimental procedure is described in Appendix~\ref{apdx:MD}) is carried out for all generated molecule graphs to return the set of MD simulated conformations. Then the coverage metric and matching metric denoted as Cov and Mat, are computed to measure the quality of generated conformations. The precise definitions are given as follows.

\begin{equation*}
\begin{aligned}
\mathrm{COV}\left(C, \mathcal{C^*} \right)
&=\frac{1}{\left|\mathcal{C^*}\right|}\left\{\sum _{C^* \in \mathcal{C^*}}{\mathbbm{1}( \mathrm{RMSD}(C, C^*) \leq \delta)} \right\}, \\
\mathrm{MAT}\left(C, \mathcal{C^*}\right)&= 
\min _{C^* \in \mathcal{C^*}} \mathrm{RMSD}(C, C^*),
\end{aligned}
\end{equation*}

where $C$ denotes the generated conformation, $\mathcal{C^*}$ represents the ground truth set of conformations sampled with MD simulation, $C^*$ denotes a ground truth instance from $\mathcal{C^*}$, $\mathbbm{1}(\cdot)$ is the indicator function which evaluates to 1 when the input is true otherwise 0, $\delta$ is the similarity threshold, which is set to 2\angstrom~in practice. From the above definition, the coverage metric describes the rate of ground truth conformations that are similar to the generated conformation, reflecting how likely the generated molecule is in a low energy state. The matching value is the minimum RMSD value between the generated molecule and the conformations in the ground truth set, indicating the similarity between the generated molecule and ground-truth conformations. In our experiments, we compute the coverage and matching metrics on both atom and fragment coordinates, to measure conformations quality at different levels.

 \begin{table}[t]
\caption{Evaluation results of generated conformations where 'atom' and 'frag' indicates the atom and fragment level coordinate metrics used for comparison.}
\label{tab:rmsd}

\begin{center}
\begin{tabular}{l|cccc}
\toprule
 ~ & \makecell[c]{COV$\uparrow$\\(atom)} & \makecell[c]{MAT$\downarrow$\\(atom)} & \makecell[c]{COV$\uparrow$\\(frag)} & \makecell[c]{MAT$\downarrow$\\(frag)}  \\
 \midrule
        EDM & 0.489 & 1.349 & 0.097 & 3.234 \\ 
        \method-E & \bf0.546 & \bf1.121 & 0.153 & 2.583\\
        \method-P &0.490 & 1.166 & \bf0.202 & \bf2.431\\
        \midrule[0.5pt]
        GEOM$_{\text{DRUG}}$ &0.589 & 0.494 & 0.435 & 1.494 \\
        \bottomrule
\end{tabular}
\end{center}
\end{table}

\subsubsection{Experimental Results} 

The experimental results listed in Table \ref{tab:rmsd} show that \method outperforms EDM consistently on all the evaluation metrics. Though \method only generates center coordinates in the coarse phase, it is able to achieve impressive results on both levels, indicating the great power of HierDiff in capturing both global and local geometric information. 
 
 We also demonstrate some visualization results in Fig.~\ref{vis_table}. We can see that the structures generated from EDM are more chaotic, with clear distorted rings and unexpected broken substructures. On the contrary, our model generates much more stable molecular scaffolds, by utilizing the coarse-to-fine approach. We also provide force field based experiments on sampled conformation to prove that our model generates more low-energy conformations, shown in Appendix~\ref{appd:E} for space limitation. We discuss the choice of high-level features in Appendix~\ref{appd:dif_feat}. 

\section{Conclusion}
This paper is concerned with 3D molecule generation. To address the irrational molecule structure problems, a hierarchical diffusion probabilistic model is proposed. We carefully design our method so that it can solve the combinatorially constrained structure generation problem introduced by non-autoregressive fragment generation modeling. To our knowledge, our work is the first attempt to get the best of both the globalization of the non-autoregressive model and the effectiveness of the fragment-based generation. \method generates better drug-like molecules, in terms of several widely used evaluation metrics. We believe that the proposed framework could inspire general solutions for other constrained structure generation tasks, such as protein alignment

\section*{Acknowledgements}
This work was supported by the National Key R\&D Program of China (No. 2021YFF1201600), Tsinghua University (NO.20221080053), Vanke Special Fund for Public Health and Health Discipline Development, and Beijing Academy of Artificial Intelligence(BAAI). Minkai Xu thanks the generous support of Sequoia Capital Stanford Graduate Fellowship. 

\nocite{langley00}

\bibliography{example_paper}
\bibliographystyle{icml2023}

\newpage
\appendix
\onecolumn

\section{Supplemented Details for the Methods}

\subsection{Implementation for Fragmentizing the Molecule}
To decrease the freedom in modeling large-size molecules, many models adopt fragment-based generation instead of building a model directly on atoms~\citep{yang2022molecule}. A number of methods are developed to break a molecule into a set of fragments. A good decomposing algorithm should satisfy that the derived fragment vocabulary needs to cover most of the molecular structures and also maintains a reasonable vocabulary size.

JT-VAE~\citep{jin2018junction} is the first deep-learning method that generates molecule graphs at the fragment level. It derived fragments by applying the minimum spanning tree algorithm to keep all chemical bond information while avoiding cycles. JT-VAE~\citep{jin2018junction} succeed to cover all buyable structures with a vocabulary size of less than 800. Recent works like MARS~\citep{xie2021mars}, FREED~\citep{NEURIPS2021_41da609c}, MIMOSA~\citep{fu2021mimosa}, FragSBDD~\citep{powers2022fragment} though applied different criterion on breaking bonds to generate fragment vocabulary, fragments of low frequency need to be removed from the vocabulary to keep the vocabulary size reasonable.

The chemical space of drug-like molecules is enormous. Leaving out fragments of low frequency is undesirable. Therefore, we adopt the tree decomposition algorithm from~\cite{jin2018junction} in a 3D space. The procedure of processing the molecules into fragment graphs is a four-step process. \textbf{Extract components}~We extract the set of chemical bonds which do not belongs to any rings and the set of simple rings which only represent a single topological cycle from the molecules.~\textbf{Merging}~The Bridged ring is a cluster of important chemical structures. They possess uncommon 3D conformation. Therefore, all pairs of rings are merged if the ring pair has more than two overlapping atoms.~\textbf{Edge linking}~Cycles in the fragment graph will cause problematic modeling since the decomposition for a molecule is not unique. To avoid cycles, the intersecting atom which connects more than 3 bonds is added to the graph as a fragment. Edges are linked between all fragment pairs that have overlapping atoms. The minimum spanning tree algorithm is run on this graph to remove overlapping edges.~\textbf{3D coarse set}~At last, we assign 3D geometric information using the center of mass of the atoms within the fragment and coarse features for each fragment.

However, there are still other methods that can be applied for the fragmentation of molecules. Hence, an experiment is conducted to prove that our fragmentation strategy has advantages and the results are show in Appendix~\ref{appd:frag}

\subsection{Implementation of Coarse-Grained Fragment Diffusion Model}
\label{appd:ddpm}
In this section, we describe the non-autoregressive high-level feature generative model and its likelihood computation. Though diffusion models have been receiving outstanding results in computer vision~\citep{DDPM, ho2020denoising, vahdat2021score}, it was nontrivial to apply directly on molecule fragment graphs. The graph features include integer features, continuous features, and continuous coordinates. These different vectors require different likelihood computations\cite{xu2022geodiff, edm}.

The diffusion model adds noise sequentially to the feature and coordinates like Eq.~\ref{0}. At the time $t$, the data distribution of invariant features is expected to approximate the prior distribution $\mathcal{N}(\mathbf{0}, \mathbf{I})$. However, in order to guarantee equivariance, the prior distribution for coordinates needs to be 
 invariance in SE(3) group. It has been proven that when the prior distribution is invariant and the transformations are equivariant the diffusion model estimates a SE(3)-invariant data distribution~\citep{xu2022geodiff}:
\begin{equation}
    \begin{aligned}
p_{\theta}\left(T_{g}\left(x_{0}\right)\right) & =\int p\left(T_{g}\left(x_{T}\right)\right) p_{\theta}\left(T_{g}\left(x_{0: T-1}\right) \mid T_{g}\left(x_{T}\right)\right) \mathrm{d} \boldsymbol{x}_{1: T} \\
& =\int p\left(T_{g}\left(x_{T}\right)\right) \Pi_{t=1}^{T} p_{\theta}\left(T_{g}\left(x_{t-1}\right) \mid T_{g}\left(x_{t}\right)\right) \mathrm{d} \boldsymbol{x}_{1: T} \\
& =\int p\left(x_{T}\right) \Pi_{t=1}^{T} p_{\theta}\left(T_{g}\left(x_{t-1}\right) \mid T_{g}\left(x_{t}\right)\right) \mathrm{d} \boldsymbol{x}_{1: T} \quad \text { (invariant prior } p\left(x_{T}\right)) \\
& =\int p\left(x_{T}\right) \Pi_{t=1}^{T} p_{\theta}\left(x_{t-1} \mid x_{t}\right) \mathrm{d} \boldsymbol{x}_{1: T} \quad \text { (equivariant kernels } p\left(x_{t-1} \mid x_{t}\right)) \\
& =\int p\left(x_{T}\right) p_{\theta}\left(x_{0: T-1} \mid x_{T}\right) \mathrm{d} \boldsymbol{x}_{1: T} \\
& =p_{\theta}\left(x_{0}\right)
\end{aligned}
\end{equation}
 As a result, we move the prior distribution for coordinates to a linear subspace where $\sum_i^{i=3} H_{pi}=\mathbf{0}$

The model minimizes the lower bound of the log-likelihood:
\begin{equation}\log P(H) \geq \mathcal{L}_0+\mathcal{L}_{\text {base }}+\sum_{t=1}^T \mathcal{L}_t\end{equation}
where:
\begin{align}
\mathcal{L}_0&=\log P\left(H \mid \boldsymbol{x}_0\right)\\
\mathcal{L}_{\text {base }}&=-\mathrm{KL}\left(q\left(\boldsymbol{x}_T \mid H \right) \mid P\left(\boldsymbol{x}_T\right)\right)\\
\mathcal{L}_t&=-\mathrm{KL}\left(q\left(\boldsymbol{x}_s \mid H, \boldsymbol{x}_t\right) \mid P\left(\boldsymbol{x}_s \mid \boldsymbol{x}_t\right)\right)
\end{align}
$\mathcal{L}_t$ and $\mathcal{L}_{base}$ can be computed easily by estimating the KL divergence between the estimated distribution and the target distribution. However, $\mathcal{L}_0$ needs special treatment. Following the previous works~\citep{edm, argmaxflow}, we define the $\mathcal{L}_0$ as follows:
\begin{equation}
\vspace{-15pt}
\begin{aligned}
    P\left(H_{f}^{int} \mid \boldsymbol{x}_0^{(H)}\right)&=\int_{H_{f}^{int}-\frac{1}{2}}^{H_{f}^{int} + \frac { 1 } { 2 }} \mathcal{N}\left(\boldsymbol{u} \mid \boldsymbol{x}_0^{(H_{f}^{int})}\sigma_0\right) \mathrm{d} \boldsymbol{u} \\
    P\left(H_{f}^{cont} \mid \boldsymbol{x}_0\right)&=\mathcal{N}\left(H_{f}^{cont} \mid \boldsymbol{x}_0^{(H_{f}^{cont})} / \alpha_0-\sigma_0 / \alpha_0 \hat{\boldsymbol{\epsilon}}_0, \sigma_0^2 / \alpha_0^2 \mathbf{I}\right) \\
    P\left(H_{p} \mid \boldsymbol{x}_0\right)&=\mathcal{N}\left(H_{f}^{cont} \mid \boldsymbol{x}_0^{(H_{p})} / \alpha_0-\sigma_0 / \alpha_0 \hat{\boldsymbol{\epsilon}}_0, \sigma_0^2 / \alpha_0^2 \mathbf{I}\right)
\end{aligned}
\end{equation}
For integer features, we centered the distribution to $h_{int}$ and integrate from $-1/2$ to $1/2$. While for continuous features and coordinates, the variance of the distribution is still approximated by the network. During sampling, our model used a regular reverse diffusion to generate features and coordinates. The only difference is that the integer feature dimensions are normalized using the round function.
\begin{table}[h]
\begin{center}
\vspace{-10pt}
\caption{Property-Based high-level feature}\label{tab:property_based}
\begin{tabular}{lll}
  Property & Description & Type \\ \hline
  HBA & Numbers of hydrogen bond acceptor & integer  \\
  HBD & Numbers of hydrogen bond donor & integer \\
  Charge & Numbers of explicit electric charge & integer\\
  Aromaticity & The size of aromatic ring & integer\\
  Alicyclicity & The size of alicyclic ring & integer\\
  Radius & The radius of the force filed optimized conformation & continuous\\
  PSA & Polar surface area contribution to the conformation & continuous\\
  ASA & Accessible surface area contribution to the conformation & continuous\\
\hline
\end{tabular}
\end{center}
\vspace{-20pt}
\end{table}
\begin{table}[h]
\vspace{-10pt}
\caption{Element-Based high-level feature}\label{tab:element_based}
\begin{center}
\begin{tabular}{lll}
  Property & Description & Type \\ \hline
  Hydrophobicity & Numbers of {C} element & integer  \\
  Hydrogen Bond Center & Numbers of {O, N, S, P} element & integer \\
  Negative Charge Center & Numbers of  {F, Cl, Br, I} element & integer\\
\hline
\end{tabular}
\end{center}
\vspace{-10pt}
\end{table}

\subsection{Implementation of Equivariant Neural Network}
\label{appd:EGNN}
\textbf{Improved EGNN} In the node/edge sampling process, our nodes are endowed with a set of invariant features and equivariant coordinates. Inspired by the recent equivariant neural networks~\citep{DBLP:journals/corr/abs-1802-08219, https://doi.org/10.48550/arxiv.2110.02905}, we propose an improved version of EGNN~\citep{EGNN}. Each layer is formulated as:
\begin{align*}
\vspace{-20pt}
m_{u v}&=\phi_m\left(n_u^l, n_v^l,\left\|x_u^l-x_v^l\right\|^2, \mathbf{e_{u v}^l}\right)
\\
x_u^{l+1}&=x_u^l+c \tanh \left(\sum_{v \in \mathcal{N}(u)}\left(x_u^l-x_v^l\right) \phi_x\left(m_{u v}\right)\right)\\
n_u^{l+1}&=\phi_n\left(n_u^l, \sum_{v \in \mathcal{N}(u)}\left(m_{u v}\right)\right)\\
\mathbf{e_{u v}^{l+1}}&=\phi_e\left(\mathbf{e_{u v}^l}, m_{u v},\left\|x_u^l-x_v^l\right\|^2\right)
\vspace{-30pt}
\end{align*}
$n$ and $e$ stands for the node/ edge embedding, while $c$ is a distance constant. $x$ stands for node coordination. All $\phi$ are classic MLPs. Previous works explore various kinds of techniques to maintain the equivariance of node features, however, the edge features are always ignored to encode into the latent variables. It has been proven that including edge feature updated in neural networks help improve performance~\citep{diao2022relational, zhou2022uni}. Edge latent variables are also needed for edge prediction in our methods. As a result, instead of carrying out the message passing on fully connected graphs with unified edges, we assigned edge features for sampling tasks. $\phi_{focal} $, $\phi_{edge} $, $\phi_{node} $ uses this improved network for message passing.

\subsection{Algorithm for Training and Sampling of fragments}
\renewcommand{\algorithmicrequire}{\textbf{Input:}}
\renewcommand{\algorithmicensure}{\textbf{Output:}}
\label{appd:decode}
\begin{breakablealgorithm}
    \caption{Training Algorithm for Node/edge decoding}
    \begin{algorithmic}[1]
    \Require{3D molecules set: $\{G\}$, EGNN networks: $\phi_{\text{focal}}, \phi_{\text{edge}}, \phi_{\text{node}}, \phi_{\text{refine}}$}
    \Ensure{EGNN networks: $\phi_{\text{focal}}, \phi_{\text{edge}}, \phi_{\text{node}}, \phi_{\text{refine}}$}
    \Function{C}{S, E: subgraph}
        \State $\text{feat} \leftarrow \text{Coarse-grained feature of~} n, n\in S$
        \State $\text{coord} \leftarrow \text{Position of~} n, n\in S$
        \State \textbf{return} feat, coord
    \EndFunction

    \Function{F}{S, E: subgraph}
        \State $\text{feat} \leftarrow \text{Fine-grained feature of~} n, n\in S$
        \State $\text{coord} \leftarrow \text{Position of~} n, n\in S$
        \State $\text{edge} \leftarrow \{i, j\}, \{i, j\} \in E$
        \State \textbf{return} feat, coord, edge
    \EndFunction

    \Function{frag}{n: node}
        \State $\text{feat} \leftarrow \text{Fine-grained feature of~} n$
        \State \textbf{return} feat
    \EndFunction
        
    \For {G in \{G\}}
        \State $T \sim {T \in G}$, s.t. $T$ is connected subgraph
        \State $n \sim {n \in T}$, s.t. $n$ is leaf node
        \State $m \sim {m \in T}$, s.t. $m$ is single node
        \State $\tilde{T} = T \setminus n$, $V = G \setminus T$
        \State $\hat{T} = T \setminus m$
        \State $\text{context} = (\Call{F}{\tilde{T}}, \Call{C}{V 
    \cup n})$
        \State $\mathcal{L}_{\text{sample}} = - \log P_{\phi_{\text{focal}}} (n.\text{parent} \mid \text{context})$ \
        
                                ~~~~~~~~~~~~~~$- \log P_{\phi_{\text{edge}}} (\{n, n.\text{parent}\}\mid 
 \text{context})$\ 
                                
                                ~~~~~~~~~~~~~~$- \log P_{\phi_{\text{node}}} (\Call{Frag}{n} \mid  \text{context}, \{n, n.\text{parent}\})$
        \State Update $\phi_{\text{focal}}, \phi_{\text{edge}}, \phi_{\text{node}} \leftarrow 
        \text{Optimize}(\mathcal{L}_{\text{sample}})$
        \State $\mathcal{L}_{\text{refine}} = -\log P_{\phi_{\text{refine}}} (\Call{Frag}{m} \mid[(\Call{F}{\hat{T}}, \Call{C}{m}])$
        \State Update $\phi_{\text{refine}} \leftarrow \text{Optimize}(\mathcal{L}_{\text{refine}})$
    \EndFor

    \end{algorithmic}
\end{breakablealgorithm}

\begin{breakablealgorithm}
    \caption{Sampling Algorithm for Node/edge decoding}
    \begin{algorithmic}[1]
    \Require{Nodes with coarse-grained feature and positions: $N$ \

            Refine step limit: max steps, \
             
            EGNN networks: $\phi_{\text{focal}}, \phi_{\text{edge}}, \phi_{\text{node}}, \phi_{\text{refine}}$}
    \Ensure{Fine-grained Graph $T$}
    \Function{Generate Step}{T}
        \State $n_{\text{focal}} \sim P_{\phi_{\text{focal}}}(n_{\text{focal}} \mid T)$
        \State $\{n_{\text{focal}}, n_{\text{new}}\} \sim P_{\phi_{\text{edge}}}(\{n_{\text{focal}}, n_{\text{new}}\} \mid T)$
        \State $T \leftarrow T + \{n_{\text{focal}}, n_{\text{new}}\}$
        \State $\text{fragment} \sim P_{\phi_{\text{node}}}(fragment \mid T)$
        \State $\Call{frag}{n_{\text{new}}} \leftarrow \text{fragment}$
    \EndFunction

    \Function{Refine Step}{T}
        \State $\text{T}_{\text{coarse}} \leftarrow$ coarse-grained T
        \State $\text{T}_{\text{fine}} \leftarrow$ fine-grained T
        \State $n_{\text{refine}} = \arg\min_{n} (P_{\phi_{\text{refine}}}(\Call{frag}{n} \mid \text{T}_{\text{fine}} \setminus n, \text{T}_{\text{coarse}}), n \in T)$
        \State $\text{fragment} \sim P_{\phi_{\text{refine}}}(\text{fragment} \mid \text{T}_{\text{fine}} \setminus n, \text{T}_{\text{coarse}})$
        \State $\Call{frag}{n_{\text{refine}}} \leftarrow \text{fragment}$
    \EndFunction

    \Function{Prob}{T}
        \State{ return $\sum_{n \in T} (\log P_{\phi_{\text{refine}}}(\Call{frag}{n} \mid T \setminus n))$}
    \EndFunction
    
    \State{$T \leftarrow N$}
    \Repeat{}
        \State{$T \leftarrow$ \Call{Generate Step}{$T$}}
        \For{i in max steps}
            \State{$\hat{T} \leftarrow$ \Call{Refine Step}{$T$}}
            \If{\Call{Prob}{$\hat{T}$} $>$ \Call{Prob}{$T$}}
                \State Accept: $T \leftarrow \hat{T}$
            \Else
                \State \textbf{Break}
            \EndIf
        \EndFor

    \Until{$\forall_{ n \in T}, n \text{~is fine-grained node}$}
    
    \end{algorithmic}
\end{breakablealgorithm}

\subsection{Algorithm for Atom-level Conformation Sampling}
\label{appd:alg-f2a}
\begin{breakablealgorithm}
    \caption{Algorithm for Conformation Alignment}
    \begin{algorithmic}[1]
    \Require{Fragment center coordinate: $F_{out}$, Molecule fragment graph: $G$}
    \Ensure{ $C_{out}$}
    \Function{Kabsch}{$X \in \mathbb{R}_3, \hat{X} \in \mathbb{R}_3$} 
        \State $X_{c} = \sum_{i=1}^n X_i, \hat{X}_{c} = \sum_{i=1}^n \hat{X}_i$
        \State $X = X - X_c, \hat{X} = \hat{X} - \hat{X}_c$ 
        
        \State $H=\sum_{i=1}^n X\hat{X}^T$
        \State $H=U \Lambda V^T$
        \State $R = {\left( UV^T \right)}^T$
        \State $t = \hat{X}_{c} - R X_{c}$
        \State \textbf{return} $R, t$ 
    \EndFunction
    
    \State $C_{\text{in}}, F_{\text{in}} \leftarrow$ RDkit random conformation and fragment positions
    \State $C_{\text{out}} \leftarrow C_{\text{in}}$
    
    \For {$n \in BFS(G)$}
        \State $n_{\text{frag}}, n_{\text{atom}} \leftarrow \text{fragment index , atom index of~}n$
        \State $n_{\text{frag}}^{\text{nei}}, n_{\text{atom}}^{\text{nei}} \leftarrow \text{fragment index , atom index of~}n.\text{neighbors}$
        \If {$n$ is not root}
            \State ${n}_{\text{frag}}^{\text{par}}, {n}_{\text{atom}}^{\text{par}} \leftarrow \text{fragment index , atom index of~}n.\text{parent}$
            \State $n_{\text{attach}} \leftarrow {n}_{\text{atom}}^{\text{par}} \cap n_{\text{atom}}$
            \State $\text{ref} = \left\{F_{\text{in}}\right[n_{\text{frag}}, n_{\text{frag}}^{\text{nei}} \left], C_{\text{in}}\right[n_{\text{attach}}\left] \right\}$
            \State $\text{out} = \left\{F_{\text{out}}\right[n_{\text{frag}}, n_{\text{frag}}^{\text{nei}} \left], C_{\text{out}}\right[n_{\text{attach}}\left] \right\}$
        \Else
            \State $\text{ref} = \left\{F_{\text{in}}\right[n_{\text{frag}}, n_{\text{frag}}^{\text{nei}} \left] \right\}$
            \State $\text{out} = \left\{F_{\text{out}}\right[n_{\text{frag}}, n_{\text{frag}}^{\text{nei}} \left] \right\}$
        \EndIf
            \State $R, t = \Call{KABSCH}{\text{ref}, \text{out}}$
            \State $C_{\text{out}}\left[n_{\text{atom}}\right] = R C_{\text{out}}\left[n_{\text{atom}}\right] + t$
    \EndFor

    \end{algorithmic}
\end{breakablealgorithm}

\section{Discussion on potential negative societal impact}
Incorporating 3D information into molecule generation could have the potential to bring a significant impact on the drug discovery industry. One key limitation of applying machine learning/ generative model approaches to the area lies in the data bias, i.e., how the 3D conformation is obtained. Such limitations could make the model difficult to generalize in practical applications. Besides, to involve human feedback, it would be ideal for the model to be model interpretable. It could be hard to obtain interpretable knowledge from the proposed generative model. The safety issues should be well taken care of towards AI-guided drug discovery. The generalization ability of such methods is a lack of exploration. More consistent and comprehensive tests are needed before the clinical tests.

\section{Additional Experiments}
\subsection{Experimental configuration}
\label{apdx:exp}
Both our model and the baseline model are trained on the GEOM$_{\text{DRUG}}$~\citep{geom}, CrossDocked2020~\citep{crossdock} and QM9~\citep{geom}. In  GEOM$_{\text{DRUG}}$ experiments, we randomly selected 4 conformations of each molecule to train our model. To test EDM~\citep{edm}, we removed hydrogen atoms from the conformations and retrained the EDM model. The implicit hydrogen atoms are reconstructed using RDkit after all other heavy atoms are generated. Because EDM only generates atom types and coordinates, a proportion of sampled molecules are not fully connected. The broken fragments were removed for numerical evaluation. In all non-autoregressive methods, the number of nodes used for sampling is drawn from the size distribution histogram calculated on the training set.

\label{apdx:MD}
\textbf{Conformation Generation}~~In this paragraph, we introduce the way to generate ground truth conformation using MD simulation. Firstly, 50 initial conformations are generated for each molecule graph using RDkit and optimized by MMF field. Then, these conformations are further optimized by MD software XTB, while the energy terms are computed for each conformation. At last, we choose the conformation with the minimum energy to sample the ground truth conformations using MD software CREST. To balance efficiency and accuracy, we set the level of optimization to 'normal' in the software for both energy computing and conformation sampling. It took approximately 16 days to generate conformations for 400 different molecules on a 128-core server.


\subsection{Stability and Validity Evaluation}
\textbf{Metrics}To illustrate our model's capacity to generate chemically valid molecule structures, we conduct experiments to compare the validity and stability with baseline models. Since the baseline EDM model is not able to generate any valid molecule with full Hydrogen coordinates on GEOM$_{\text{DRUG}}$, we compared our model with the EDM model that needs to sample Hydrogen coordinates with the help of RDkit. The mol stability is defined as the proportion of molecules that can be interpreted as valid molecules with all bonds and coordinates explicitly defined in RDkit. The validity is defined as the proportion of molecules that are connected rather than individual fragments in 3D space. Diversity measures the diversity of the generated molecules by calculating the average pairwise Tanimoto Morgan Fingerprint, following previous work~\citep{xie2021mars}.

Although our model is designed to generate drug-like molecules with relatively large molecule sizes, it can be applied for smaller organic molecule~(QM9) generation tasks without effort. We also measure the validity and uniqueness metric from previous works~\citep{edm} on 10000 generated small organic molecules and compared them with various baselines by using RDkit.

\textbf{Baselines} Our method is compared with previous methods. Both graph-based and coordinate-based models are included here. Graph-based methods like Graph VAE~\citep{simonovsky2018graphvae}, GTVAE~\citep{mitton2021graph}, and Set2GraphVAE~\citep{vignac2021top}, do not explicitly define the coordinates, so they need cheminformatic software to generate conformers. On the other hand, 3D coordinate-based models like E-NF~\citep{enflow}, G-Schnet~\citep{Gschnet}, and EDM~\citep{edm}, need cheminformatic software to derive chemical bonds. 
\begin{table}[h]
\caption{Molecule stability metrics on QM9. 3D: model generation in 3D space. Bond: model chemical bonds.}\label{tab:qm9}
\begin{center}
\begin{tabular}{lcccccc}
  Method & 3D & Bond & Valid~(\%) & Unique~(\%)\\ \hline
  GraphVAE &  ~ & \checkmark & 55.7 & 75.9 \\ 
  GTVAE & ~ &  \checkmark & 74.6 & 22.5 \\
  Set2Graph &  ~ & \checkmark & 59.9 & 93.8 \\
  E-NF & \checkmark  & ~ & 40.2 & 98.0 \\
  G-Schnet & \checkmark  & ~ & 85.5 & 93.9 \\
  EDM & \checkmark  & ~ & 91.9 & 98.7 \\
  \method$_\text{E}$~(ours)& \checkmark  & \checkmark & 87.8 & 97.9\\
  \method$_\text{P}$~(ours)& \checkmark  & \checkmark & 83.6 & 98.5\\
\hline
\end{tabular}
\end{center}
\end{table}

\begin{table}[h]
\caption{Molecule stability metrics on GEOM$_{\text{DRUG}}$}\label{tab:geom_stab}
\begin{center}
\begin{tabular}{lcccc}
  Method & Mol Stability & Unique & Validity & Diversity\\ \hline
  EDM~(without H) & 0.970 & 1.000 & 0.835 & 0.824 \\ 
  \method$_\text{E}$~(ours)& 1.000  & 1.000 & 0.940 & 0.836 \\
  \method$_\text{P}$~(ours)& 1.000  & 0.995 & 0.904 & 0.832\\
\hline
\end{tabular}
\end{center}
\end{table}

\textbf{Results} When testing the stability metrics on GEOM$_{\text{DRUG}}$, our methods outperform EDM in stability, validity, and diversity. This indicates that our model generates not only accurate conformations of drug-like molecules but also enjoys great sampling efficiency. As shown in Table~\ref{tab:qm9}, our method performs comparable results in both validity and uniqueness. Though EDM~\citep{edm} achieved better performance, our method still outperforms all other models. The slight performance drop compared to EDM could be due to the information loss ratio during fragmentization on the tiny graphs.  Besides, our model is the only 3D method that does not depend on any chemical bond linking software, like Openbabel. Hydrogen atoms can be added by counting the valency for each atom in our method.
\vspace{-10pt}
\subsection{Additional evaluation of drug-like properties}
\textbf{Ring Size} Ring Systems with the size of 5-6 are stable chemical groups in organic chemical theories.
\textbf{HeteroAtom} The number of heteroatoms represents area of the polar surface in the organic molecules, which highly determines the distribution of the drug molecules in the human body, \textit{e.g.}, the drug molecules that can cross the blood-brain barrier always has fewer heteroatoms.
\textbf{AromaticRing} The Number of aromatic rings in the molecules indicates the ability to form $\pi-\pi$ interaction with proteins or other biomolecules. Aromatic rings also stabilize the molecule into lower energy conformations.
\textbf{AliphaticRing} The Number of aliphatic rings in the molecules indicates the rigidity of the molecules. Instead of lying in a plane as aromatic rings, aliphatic rings constrained the conformation by contributing a specific torsion angle to the molecule conformation.
\textbf{Radius} The mean radius of the fragment. A higher radius than the GEOM$_\text{DRUG}$ indicates too many rings are generated by the model. A smaller radius indicates the model is not able to construct valid ring systems.

\begin{figure}[h]
\begin{center}
\vspace{-15pt}
\includegraphics[width=0.4\textwidth]{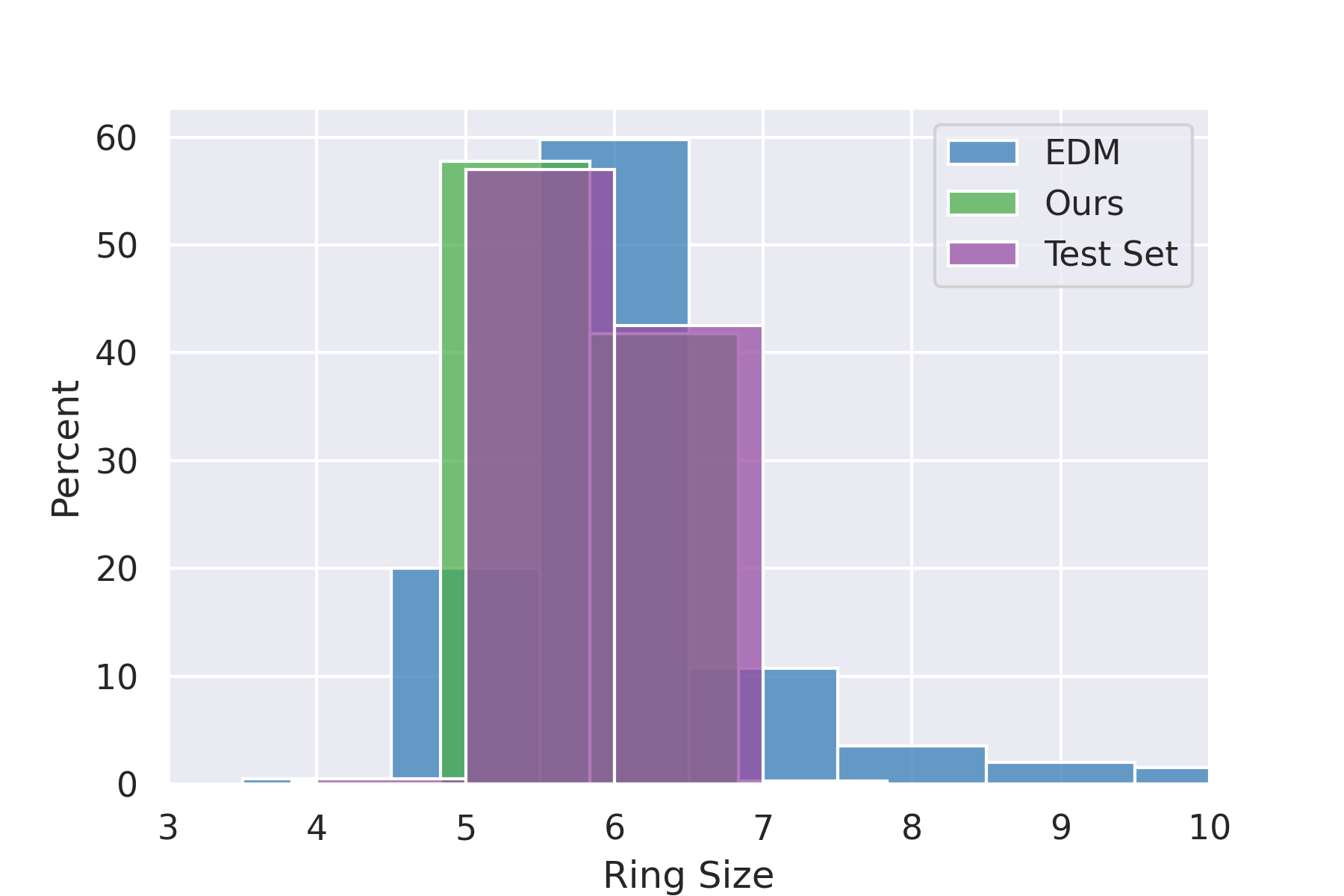}
\end{center}
\caption{Histogram of the sampled molecules' ring size frequency of our model, EDM, and GEOM$_\text{DRUG}$}
\label{fig:ring_size}
\end{figure}

\textbf{Result and Discussion} In addition to the evaluation of properties on the molecule level, we also break all the sampled molecules into fragments and test their performance on additional properties
As expected, our method chooses fragments that are similar to that of ground truth statistics.
We plotted the distribution of ring size in Fig. \ref{fig:ring_size}, the number of our method conforms best with ground truth. It is obvious that ring sizes 5 and 6 are most commonly seen in drug datasets and our sampled results, which are stable. However, on the contrary, the atom-based method such as EDM~\citep{edm} has failed to capture this basic chemical rule. 
 Refer to Table \ref{tab:frag_prop} for additional property evaluation.\\

\begin{table}[h]
\vspace{-15pt}
\caption{Properties of the fragments. All molecules are decomposed into fragments for statistical analysis. The model performs better if the generated fragments have more similar properties to GEOM$_\text{DRUG}$.}
\label{tab:frag_prop}
\begin{center}
\begin{tabular}{llllllll}
 ~ & Ring Size  & HeteroAtom  &AromaticRing &AliphaticRing &Radius \\ \hline
        EDM & 6.038 &0.605& 0.065 &0.101 &1.265  \\
        \method$_\text{E}$ & \bf5.749  & 0.630  & 0.104 &\bf0.083 & 1.295  \\ 
        \method$_\text{P}$ & 5.714 & \bf0.660 & \bf0.132 & 0.082 &\bf1.360 \\ \hline
        GEOM$_\text{DRUG}$ & 5.747  &0.677   &0.134 &0.066 &1.351  \\ \hline
\end{tabular}
\end{center}
\end{table}

\subsection{Conformation Energy Experiments}
\label{appd:E}
Besides conformation quality evaluation, we also want to verify that our model generates conformation with realistic energy terms. We compared our model with two baseline methods, EDM~\citep{edm} and JT-VAE~\citep{jin2018junction}. JT-VAE is a 2D fragment-based molecule generation model, so we used the RDkit ETKDG module to generate the 3D conformation for the sampled 2D molecules. Similar to the conformation quality experiments, we trained all models on GEOM$_\text{DRUG}$ dataset and sampled 100 conformations with each method to carry out our experiments. We compute the energy using Merck Force Field(MFF). MMD distance with the Gaussian kernel is applied here to measure the difference between generated distributions and the original distribution.

\begin{figure}[h]
\begin{center}
\includegraphics[width=0.9\textwidth]{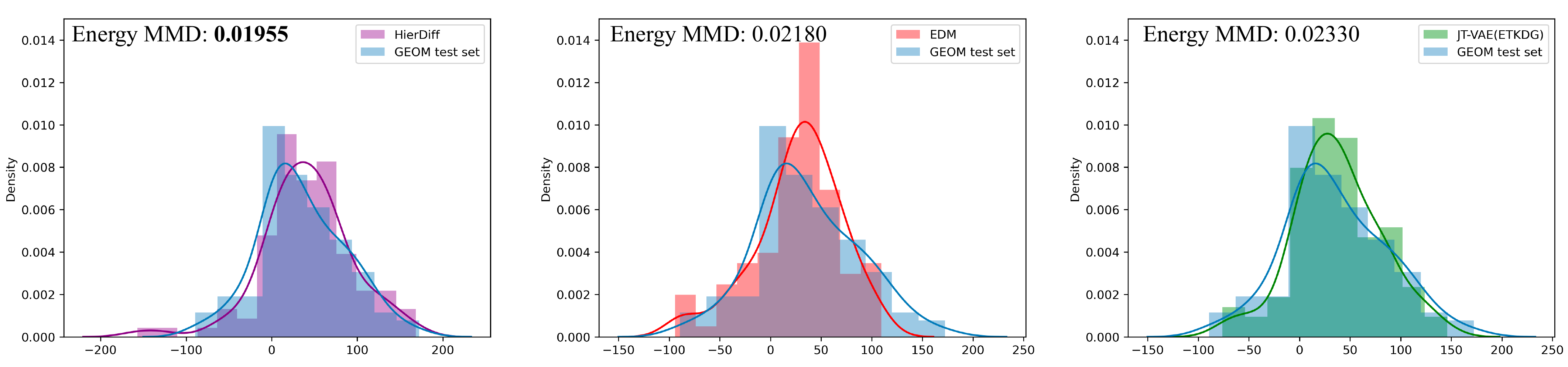}
\end{center}
\caption{MFF Energy distribution of the generated 3D molecules. Lower Energy MMD means the model is able to generate more stable conformation.}
\label{fig-energy}
\end{figure}

\textbf{Result and Disscussion} According to Fig.~\ref{fig-energy}, \method generates conformations with the closest energy distribution with the original dataset. Though EDM is outperformed by our method, it still beats JT-VAE(ETKDG) in generating more stable conformations. We can conclude that in this task, 3D methods have advantages over using RDkit conformation sampling along with a 2D method.

\subsection{Evaluation of Uniqueness and Diversity on GEOM$_{\text{DRUG}}$}
To prove that our method does not occur the issue of mode collapse, we tested the uniqueness of generated molecules and evaluate the similarity of generated molecules with the  GEOM$_{\text{DRUG}}$ test set. \textbf{Similarity} which measures the average similarity between generated molecules with the most similar molecule in the test set. We use the Tanimoto score between ECFP4 fingerprints to measure the similarity between two molecules. High similarity indicates that the method lack generalization. \textbf{Unique} is the proportion of unrepeated structures in generated molecules. Both metrics are tested on molecule level and Murko Scaffold level. Numeric results are listed in Table~\ref{tab:diversity} It is quite clear that our model generates more diverse molecules.

\textbf{Result and Disscussion}~ \method outperforms EDM~\citep{edm} on this uniqueness and diversity test. It should be noted that both our method and EDM generate mostly unique molecules when training on GEOM$_\text{DRUG}$. These methods succeed to generate diverse 3D molecules. Combining to results from Table~\ref{exp:property}, our method is able to generate 3D molecules that are drug-like and diverse.

\begin{table}[h]
\caption{Diversity metrics computed on 1000 drug-like molecules generated by our method with two types of coarse feature and EDM~\citep{edm}.}
\label{tab:diversity}
\begin{center}
\begin{tabular}{lcccc}
 ~ &  Similarity-atom~$\downarrow$ & Unique-atom~$\uparrow$  &  Similarity-scaffold~$\downarrow$ & Unique-scaffold~$\uparrow$ \\ \hline
        EDM & 0.176 & 1.000 & 0.189 & 0.930 \\
        \method$_\text{E}$ & \bf{0.164} & 1.000 & 0.171 & \bf{0.957} \\ 
        \method$_\text{P}$ & 0.168 & 1.000 & \bf{0.169} & 0.946 \\ \hline
\end{tabular}
\end{center}
\end{table}

\subsection{Conditional generation}
In the field of AI-guided drug discovery, one of the most essential directions is to generate 3D molecules according to desirable properties. Though previous works~\citep{Gschnet, edm} have tested their abilities to generate molecules conditioned on simulated energy, to our knowledge, we are the first to carry out practical drug discovery-related property-conditional generation experiments using the 3D molecule generative model. We include the properties as an additional atom feature dimension in our dual-phase generation process. We tested the difference using mean squared error~(MSE) and mean absolute error~(MAE) between the input properties and the real properties of the generated 3D molecules. The diversity of the generated molecules is also evaluated using mean fingerprint similarity as defined in \citet{xie2021mars}.

\begin{tabular}{ccccccccccccc}
\hline & & Asphericity & & & QED & & & SAS & & & $\log P$ & \\
\hline . & MSE & MAE & Div & MSE & MAE & Div & MSE & MAE & Div & MSE & MAE & Div \\
\hline EDM & 0.626 & 0.455 & \bf0.895 & \bf0.113 & \bf0.285 & 0.883 & 0.074 & 0.193 & \bf0.897 & 6.054 & 2.019 & \bf0.883 \\
 HierDiff-$_\text{E}$ & \bf0.176 & \bf0.406 & 0.894 & 0.120 & 0.289 & \bf0.885 & \bf0.051 & \bf0.184 & 0.881 & \bf1.405 & \bf0.976 & 0.882 \\
\hline
\end{tabular}

\textbf{Results and Disscusions} To make fair comparison, we excluded HierDiff-$_\text{P}$ for property-conditional generation since it already implies fragment properties. For example, aromatic rings and topological surfaces are related to logP and SAS. The results illustrate that our model outperforms EDM on Asphericity, SAS, and LogP in generating 3D molecules with accurate properties. EDM outperforms HierDiff in the accuracy of the QED conditioned generation task and the diversity in other metrics by an almost negligible margin.

\subsection{Ablation Study}
\label{appd:abl}\label{appd:frag}
We remove the iterative refinement step in each fragment sampling process when using the model training on GEOM$_\text{DRUG}$ to carry out the ablation study. The results are shown in Tab.~\ref{tab:abl}. The experiment results illustrate that the hierarchical diffusion-based model samples molecules with higher molecular weight and lower SAS score, however, the drug-likeness score QED and safety score MCF decreased.
\begin{table*}[h]
\caption{Properties of the generated molecules. $\Delta$ indicates that the evaluated metrics are computed as the difference between sampled molecules and the ground truth and the absolute values are listed in the~(). '-r' is the notation that this model sample molecule without any iterative refinement.}
\label{tab:abl}
\begin{center}
\begin{tabular}{lcccccc}
    ~ & QED$\uparrow$ & RA$\uparrow$& MCF$\uparrow$& SAS$\downarrow$ & $\Delta_{\text{LogP}}\downarrow$ ~(logP) & $\Delta_{\text{MW}}\downarrow$ ~(MW) \\ \hline
    \method$_\text{E}$~(-r) & 0.628 & 0.626 & 0.681 & 3.669 & 0.638~(2.291) & 27.33~(332.8) \\
    \method$_\text{P}$~(-r) & 0.635 & 0.638 & 0.656 & \bf{3.512} & 0.185~(2.744) & \bf10.33~(349.8) \\
    \method$_\text{E}$ & 0.632 & 0.548 & \bf0.727 & 3.859 & 0.653~(2.276) & 30.33~(329.8) \\
    \method$_\text{P}$ & \bf0.639 & \bf0.643 & 0.659 & 3.547 & \bf0.128~(2.801) & 13.33~(346.8) \\ \hline
    GEOM$_{\text{DRUG}}$ & 0.658 & 0.915 & 0.774 & 4.018 & 0.000~(2.929) & 0.000~(360.1) \\ \hline
\end{tabular}
\end{center}
\end{table*}

Besides, we also replaced the minimum spanning tree algorithm with the 'High-frequency Fragments' strategy and 'Single Rings and Bonds' strategy for an ablation study. The 'High-frequency Fragment' is a common strategy applied in previous works~\citep{xie2021mars, powers2022fragment} in which all fragments are collected from the training datasets using basic predefined rules. Only the high-frequency fragments are kept in the vocabulary in this method. 'Single Rings and Bonds' is another simple fragmentation strategy in that the fragments are limited to all single bonds and single rings. Bridge rings and multiple-ring systems are broken into single rings to reduce fragment complexity. The results are listed in Tab.~\ref{frag_compare}, which illustrate that without using a complicated fragmentation strategy our method still outperforms the baseline model in most metrics. We chose the minimum spanning tree-based strategy because it offers a more balanced performance in all drug-like properties.
\begin{table*}[h]
\caption{Properties of the generated molecules generated by different fragmentation strategies. $\Delta$ indicates that the evaluated metrics are computed as the difference between sampled molecules and the ground truth and the absolute values are listed in the~().}
\label{frag_compare}
\begin{center}
\begin{tabular}{cccccccc} 
Fragmentize Methods & QED $\uparrow$ &RA $\uparrow$ & MCF $\uparrow$ & SAS $\downarrow $& $\Delta$ Log $\downarrow$ & $ \Delta$ MW $\downarrow$ & Validity $\uparrow$ \\
\hline EDM & 0.608 & 0.441 & 0.621 & 4.054 & 0.566 & 23.71 & 0.835 \\
 Minimum Spanning Tree & \bf 0.639 & \bf 0.639 & 0.659 & \bf 3.574 & \bf 0.128 & 13.33 & 0.904 \\
 High-frequency Fragments & 0.609 & 0.632 & 0.662 & 3.638 & 0.228 & 9.80 & 0.853 \\
 Single Rings and Bonds & 0.621 & 0.600 & \bf 0.745 & 3.756 & 0.680 & \bf7.81 & \bf0.932 \\ \hline
\end{tabular}
\end{center}
\end{table*}

\subsection{Discussion on Sampling Efficiency}
\begin{table*}[h]
\begin{center}
\caption{Computational time for sampling 3D molecules using different models}\label{tab:time}
\begin{tabular}{cc}
\hline Model & Sample time per sample (s)\\
\hline
EDM (reported from \citet{edm} & 10.2\\
EDM (reproduce) & 3.15 \\
\method~(1000 steps) & 5.17\\
\method~(500 steps) & 3.74\\
\method~(250 steps) & 3.02\\ \hline
\end{tabular}
\end{center}
\end{table*}
We conduct an experiment on the computation cost(sampling speed). It is true that the decoding models $\phi_{focal}$, $\phi_{node}$ and $\phi_{edge}$ indeed bring some computational cost. While these modules are not the key bottleneck as only a few forward passes need to be conducted during sampling. As shown in Tab.~\ref{tab:time}, the diffusion phase is actually the main computational overhead as there are many forward passes, e.g. thousands of steps, that need to be conducted during a single sampling process. Besides, we also noted the fact that our model is capable of using a larger batch size for parallel on a GPU since our diffusion space is smaller. Your question actually inspires us to train our model in a new setting with fewer diffusion steps, for example, 500 steps or 250 steps. We leave this direction as future exploration.
\subsection{Disscusion on the Choice of high-level feature}
\label{appd:dif_feat}
From the above experiments, we can see that both the property-based coarse feature and the element-based coarse feature outperform the baseline models in generating more drug-like molecules and sample stable conformations. However, these two kinds of features reveal different strengths. 

In the Drug-likeness evaluation, HierDiff-P outperforms HierDiff-E in most metrics when training on GEOM$_{\text{DRUG}}$. HierDiff-E achieves the highest score on MCF. The reason is that property-based features provide more chemical semantic information than element-based features. HierDiff-E achieves better results on chemical safety because the element histogram helps the model avoid generating combinations of elements that can form unstable subgraphs. On the contrary, when training on CrossDock, HierDiff-E performs better, because CrossDock is a smaller dataset. It's easier to approximate the true distribution with a simpler representation as the latent variable.

In the conformation quality evaluation, HierDiff-E achieves better results on the atom level RMSD and HierDiff-P achieves better results on the fragment level RMSD. Since element-based features directly model on the atom level and property-based features include connection information, surface information on a global view, this result agrees with the motivation that inspires us to design these features.

\subsection{Experimental Proof of Error Accumulation}
\begin{figure}[h]
\begin{center}
\includegraphics[width=0.5\textwidth]{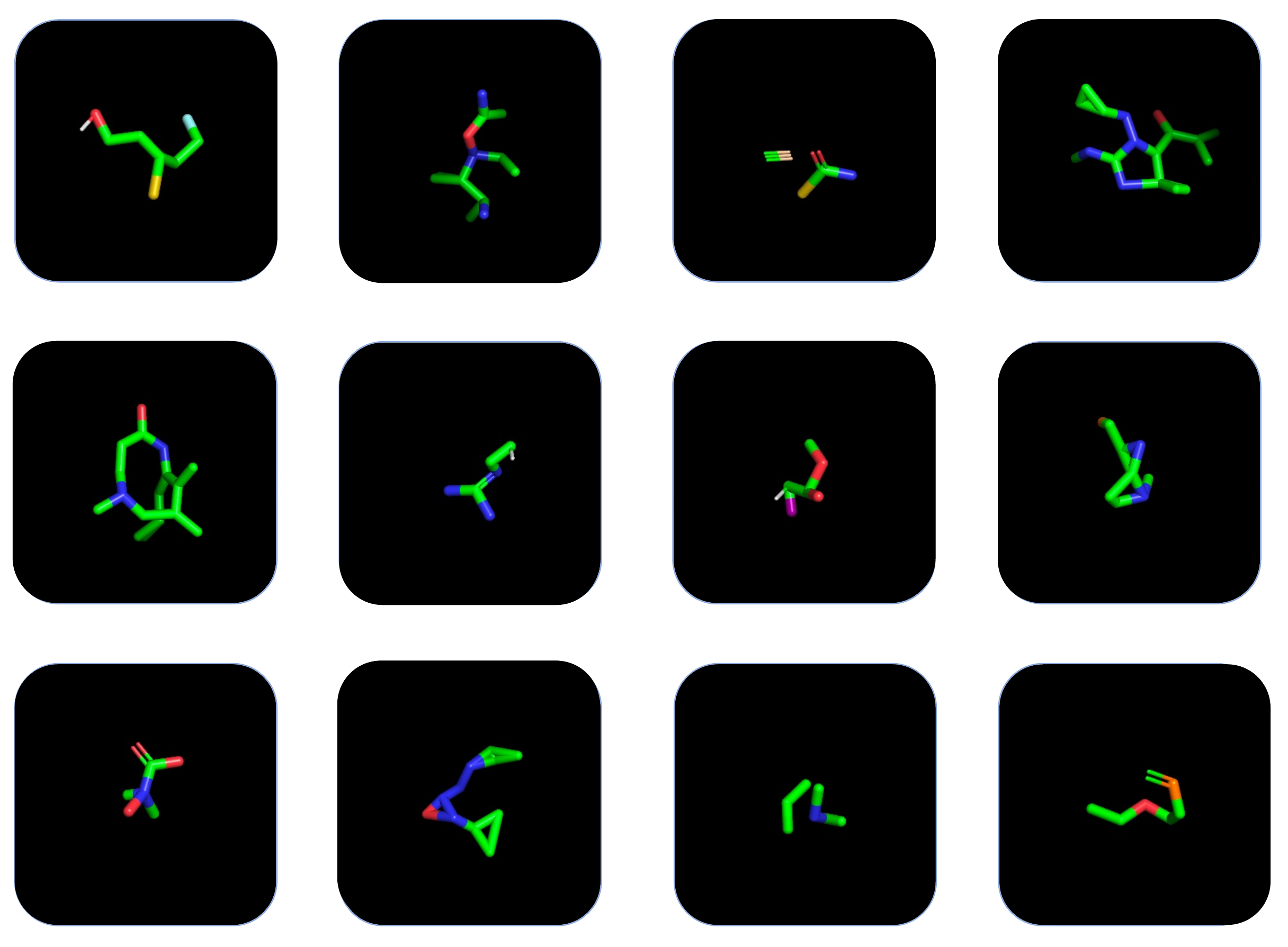}
\end{center}
\caption{Visualized 3D conformations generated by G-SphereNet}
\label{fig:sphere-case}
\end{figure}

One of our motivations for developing a hierarchical method for molecule generation is that we discovered the error accumulation in molecule generative models. This means that when the molecule size increase, the error from the previous steps of generation influences later steps which leads to unrealistic results. This issue has been discussed in the field of natural language modeling~\citep{schmidt2019generalization, he2019exposure, caccia2018language}. To prove this issue exists, we trained our method which represents the non-autoregressive method, and G-Spherenet which represents the autoregressive method on QM9. Both methods are set to generate molecules with the given molecule size. We test the validity of the generated molecules. We also do the same test on GEOM$_{\text{DRUG}}$, however, the validity of the autoregressive model drops so fast that it cannot generate molecules with more than 20 heavy atoms. Numeric results on QM9 are listed in Table~\ref{appd:error-accum}. Visualized results of  GEOM$_{\text{DRUG}}$ can be found in Figure~\ref{fig:sphere-case}.

\begin{table}[h]
\caption{Validity of sampled molecules with different sizes trained on QM9~\citep{geom}. All molecules that are broken or marked as invalid in RDkit package are regarded as invalid samples. AR stands for the autoregressive model in G-SphereNet~\citep{G-spherenet}. non-AR stands for our method.}
\label{appd:error-accum}

\begin{center}
\begin{tabular}{ccccc}
    Size & 5 & 6 & 7 & 8 \\ \hline
    Valid~(AR) & 0.710 & 0.692 & 0.690 & 0.588\\
    Valid~(non-AR)  & 1.000 & 0.950  & 0.953 & 0.991 \\ \hline
   
\end{tabular}
\end{center}

\vspace{-8pt}
\end{table}


\subsubsection{More Visualization Results}

\begin{figure}[h]
\begin{center}
\includegraphics[width=0.5\textwidth]{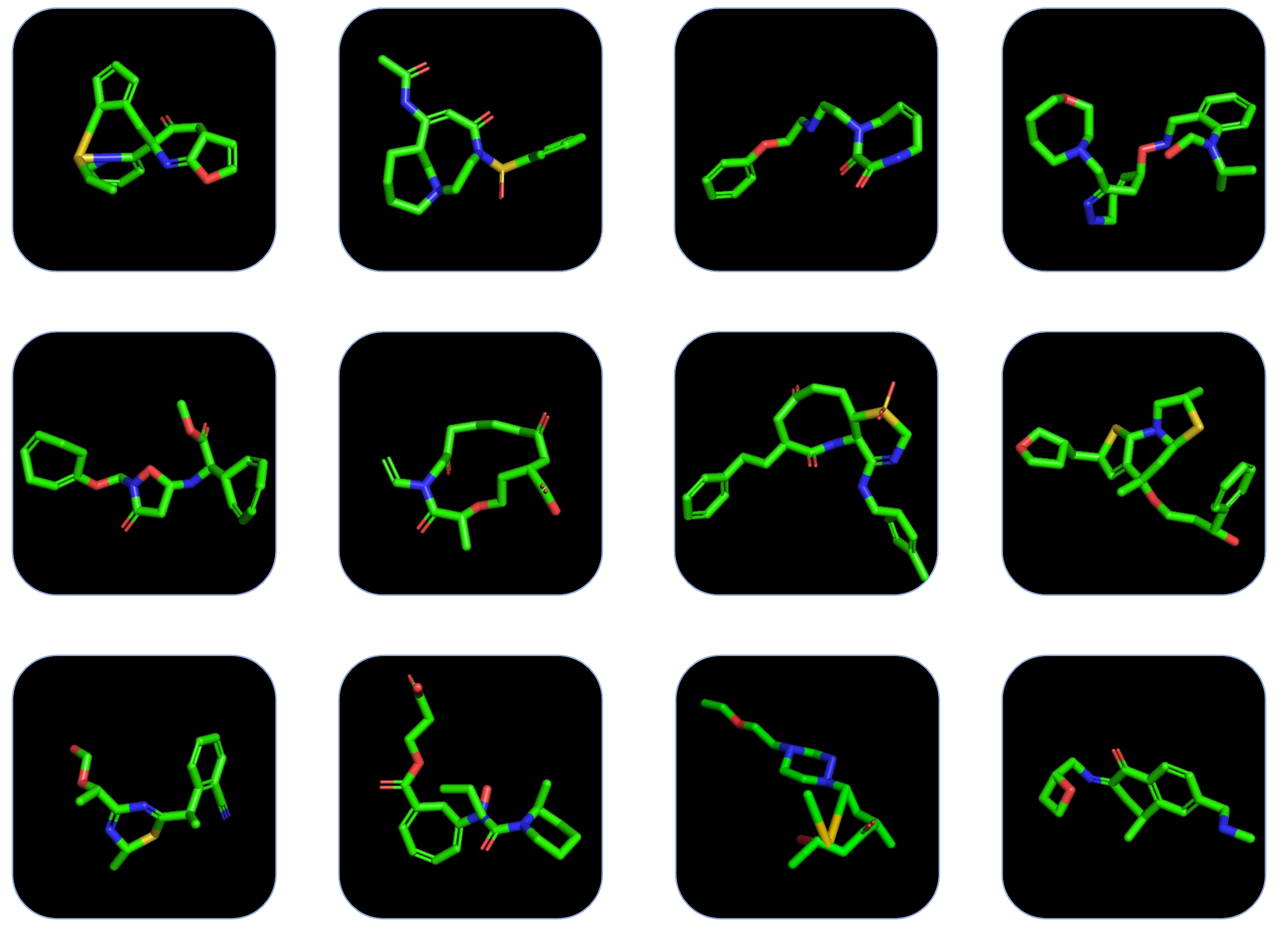}
\end{center}
\caption{Visualized 3D conformations generated by EDM}
\end{figure}

\begin{figure}[h]
\begin{center}
\includegraphics[width=0.5\textwidth]{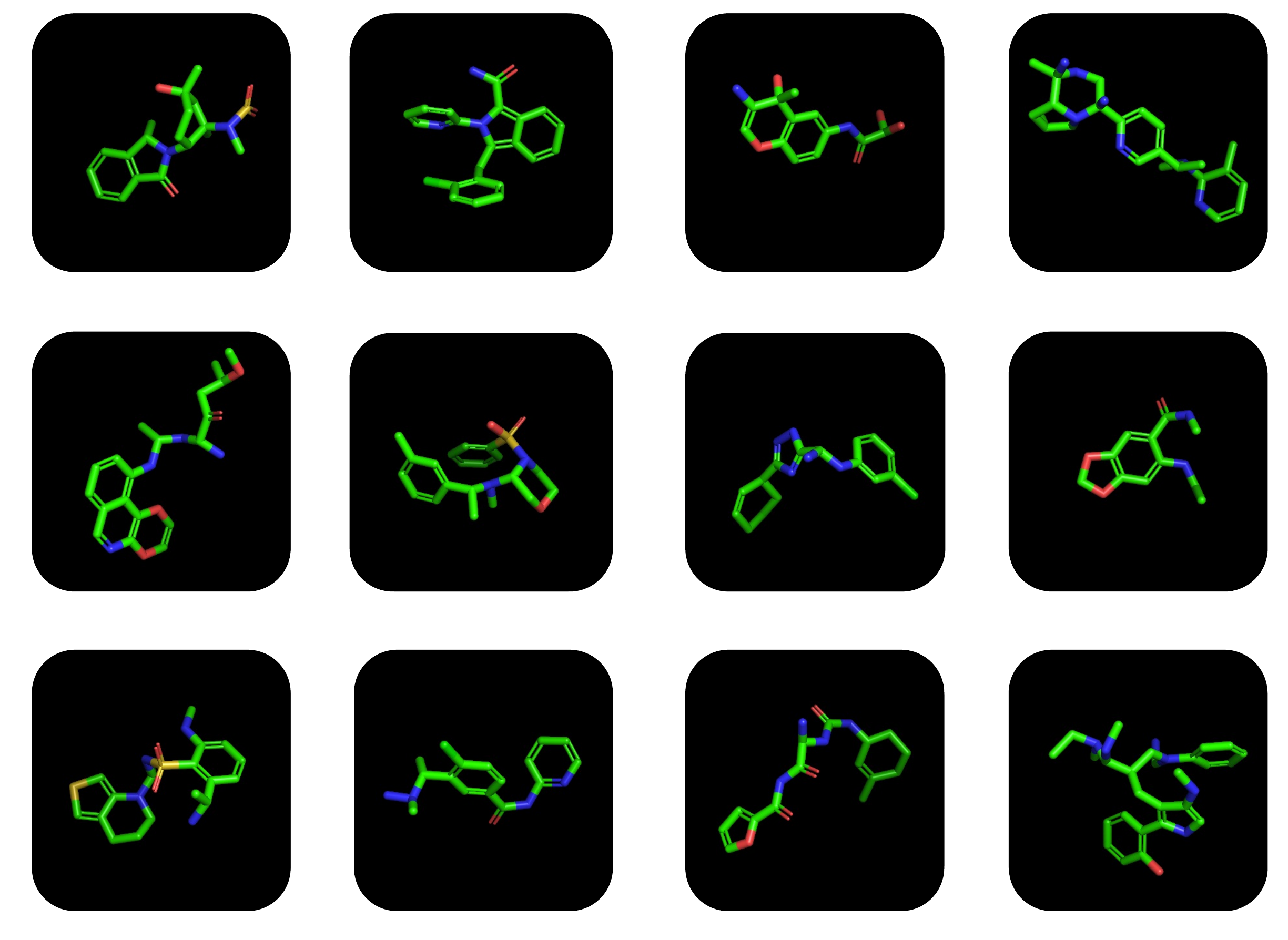}
\end{center}
\caption{Visualized 3D conformations generated by \method}
\end{figure}

\end{document}